\documentclass[12pt]{iopart}
\usepackage{iopams}
\usepackage{graphicx,subfig}
\usepackage{color}
\usepackage{bm}
\usepackage{epstopdf}
\def\(({\left(}
\def\)){\right)}
\def\[[{\left[}
\def\]]{\right]}

\newcommand{\be}{\begin{equation}}
\newcommand{\ee}{\end{equation}}
\newcommand{\bea}{\begin{eqnarray}}
\newcommand{\eea}{\end{eqnarray}}
\begin{document}
\title{Following states in temperature in the spherical $s+p$-spin glass model}

\author{ YiFan Sun$^{1,2}$, Andrea Crisanti$^{3,4}$, Florent Krzakala$^{2}$, Luca Leuzzi$^{3,5}$ and Lenka Zdeborov\'a$^6$}

\address{$^1$ LMIB and School of Mathematics and Systems Science, Beihang University, 100191 Beijing, China}
\address{$^2$ CNRS and ESPCI ParisTech, 10 rue Vauquelin, UMR 7083 Gulliver, Paris 75005, France}
\address{$^3$ Dept. of Physics, University {\em Sapienza}, Piazzale Aldo Moro 5, 00152, Rome, Italy}
\address{$^4$ ISC-CNR {\em Sapienza}, Piazzale Aldo Moro 5, 00152 Rome, Italy}
\address{$^5$ IPCF-CNR Roma {\em Kerberos}, Piazzale Aldo Moro 5, 00152 Rome, Italy}
\address{$^6$ Institut de Physique Th\'eorique, IPhT, CEA Saclay, and URA 2306, CNRS, 91191 Gif-sur-Yvette, France}

\ead{\mailto{yfsun1984@gmail.com}}
\begin{abstract}
  In many mean-field glassy systems, the low-temperature Gibbs measure
  is dominated by exponentially many metastable states.  We analyze
  the evolution of the metastable states as temperature changes
  adiabatically in the solvable case of the spherical $s+p$-spin glass
  model, extending the work of Barrat, Franz and Parisi [{\em
    J. Phys. A} {\bf 30}, 5593 (1997)]. 
  We confirm the presence of level crossings, bifurcations, and
  temperature chaos. For the states that are at equilibrium close to
  the so-called dynamical temperature $T_d$, we find, however, that
  the following state method (and the dynamical solution of the model
  as well) is intrinsically limited by the vanishing of solutions with
  non-zero overlap at low temperature.
\end{abstract}
\vspace{2pc}
\pacs{64.70.qd, 75.10.Nr, 75.50.Lk}
\noindent{\it Keywords}:
\maketitle

\section{Introduction}

Predicting the dynamical behavior of a system from a calculation of
static quantities is the basic goal of statistical mechanics, where we
consider the static average over all configurations rather than the
joint dynamics of all the atoms. A particularly important case, both
in classical and quantum thermodynamics, is given by the dynamics
after very slow variations of an external parameter so that the system
remains in equilibrium. Such changes are said to be ``adiabatic".

In mean-field systems  it takes an exponentially large time, in the
size of the system, to escape from a metastable state. It is, thus,  a well
posed question to study the ``adiabatic'' evolution of each such
metastable state. One simply needs to consider that the speed of change
of the external parameter is very slow but independent of the system
size.

In glassy mean-field systems, at low temperatures, there are
exponentially many metastable states. The equilibrium solution in
these systems can be described using the replica theory or the cavity method
\cite{MezardParisi87b}. Adiabatic evolution of the states can be described
using the Franz-Parisi potential \cite{FranzParisi95,FranzParisi97}. One fixes
a reference equilibrium configuration at temperature $T_e$ and computes the
free energy at temperature $T_a$ restricted to configurations sharing a given degree of similarity with the reference configuration. A distance can be defined in terms of the overlap of system configurations with the reference configuration  and the constrained free energy  is  the Franz-Parisi potential. The adiabatic
 evolution of the metastable state, to which  the reference
 configuration belongs, can, then, be represented by the set of local minima of the
 Franz-Parisi potential at the shortest distance from this configuration.

 Krzakala and Zdeborova recently introduced a
 procedure for following states that focuses directly on the minimum
 of the Franz-Parisi potential and is more easily tractable from a
 computational point of view
 \cite{KrzakalaZdeborova09b,ZdeborovaKrzakala10}.  The method has been applied
 to a number of glassy mean field systems including the fully
 connected Ising $p$-spin glass model, the diluted Ising $p$-spin model
 (XOR-SAT)  or the random graph coloring. In all these systems, for $T_e$ 
 below the dynamic arrest temperature $T_d$
 and for $T_a$ low enough, the replica symmetric (RS) Franz-Parisi potential ceases to have a
 local minima correlated to the reference configuration. If  the Franz-Parisi potential is, rather, evaluated using
 1RSB,  the correlation with the reference configuration vanishes at a temperature $T_a$ lower than the one for the RS case, though it still vanishes.
 It was hence hypothesized that
 introducing further steps of replica symmetry breaking a physical
 solution might  eventually be found.

The current work is motivated by these findings and consists in the  study of the
adiabatic evolution of states in the spherical $s+p$-spin glass model
with different competing $p$-body interaction terms, where an
arbitrary level of replica symmetry breaking is relatively easily
tractable \cite{CrisantiSommers91,Nieuwenhuizen95,CrisantiLeuzzi04,CrisantiLeuzzi07b}, and where, moreover, the dynamical behavior can be exactly
solved \cite{CrisantiHornerSommers93,CugliandoloKurchan93,Barrat97,CrisantiLeuzzi07}, and agrees with the results of the computation of
the Franz-Parisi potential.

When following the evolution of the states in the spherical spin glass
model as temperature is lowered we find once again the loss of
correlation with the reference configuration. In this case, though, we
are able to explicitly check that further levels of replica symmetry
breaking do not preserve the correlation from vanishing.  We will
discuss the implications of such property throughout the paper and
their possible connections to systems with discrete variables.

The rest of the paper is organized as follows. Section~\ref{model} describes the
model we study in this paper and summarizes the static replica
equations under different levels of replica symmetry breaking. Section~\ref{following} derives
the equations for evolution of states with the aid of the results we
obtain in fully connected Ising $p$-spin model.  In Section~\ref{results}, we report
results of the state evolution, discuss their physical meaning and compare them
with existing results. Finally, in Section~\ref{KZ_instability}, we discuss the vanishing of states under cooling at low temperature.

\section{Model and its thermodynamics}
\label{model}
In this paper we study the spherical 3+4  spin-glass model with ferromagnetic interaction 
(3+4-FM model) whose Hamiltonian is
\begin{equation}
\mathcal{H} = -\sum_{i_1<i_2 <i_3}J^{(3)}_{i_1,i_2,i_3}
                                      \sigma_{i_1}\sigma_{i_2}\sigma_{i_3}\,
-\sum_{i_1<i_2<i_3<i_4}J^{(4)}_{i_1,i_2,i_3, i_4}
                                      \sigma_{i_1}\sigma_{i_2}\sigma_{i_3}\sigma_{i_4}
\label{Hamiltonian-34}
\end{equation}
where the interactions $J^{(p)}_{i_1,i_2\ldots i_p}$ are i.i.d random Gaussian variables of 
mean $S_p p!/N^{p-1}$ and variance $J^2_p  p!/(2N^{p-1})$, and the 
spins $\sigma_i$ real variables satisfying the global spherical  constraint
\begin{equation}
\sum_{i=1}^N\sigma_i^2=N \, .
\end{equation}
This model is a particular case of the general spherical spin glass model with ferromagnetic
interaction defined by the Hamiltonian
\begin{equation}
\mathcal{H}=-\sum_{p=2}^M\sum_{i_1<i_2<\ldots
  <i_p}J^{(p)}_{i_1,i_2\ldots
  i_p}\sigma_{i_1}\sigma_{i_2}\ldots\sigma_{i_p}\, ,
\label{Hamiltonion}
\end{equation}
where several $p$-uples $(i_1,i_2\ldots i_p)$ of spins interacting via random
Gaussian interactions $J^{(p)}_{i_1,i_2\ldots i_p}$, with the mean and variance given above, 
are considered. 
It turns out that close to the transitions only the first two non-zero terms in the sum 
(\ref{Hamiltonion}) are relevant \cite{GotSjo89,CriCiu00}, and so one usually
considers the Hamiltonian (\ref{Hamiltonion}) where only two terms are retained.
To indicate which terms are considered these models are referred as
$s+p$-FM models. 

The equilibrium properties of these models can be obtained using the, now standard,
{\sl replica method} to average over the disorder.
 We briefly remind here the replica solution of this model as introduced in
\cite{CrisantiSommers91,Nieuwenhuizen95, CrisantiLeuzzi04,CrisantiLeuzzi07b,CriLeu12}.
Introducing replicas and 
averaging the partition function over the disorder yields  the static free energy functional, which reads
\begin{eqnarray}
-2\beta F(\beta)&=& 1+\ln 2\pi +\lim_{n\to 0}\frac{1}{n}G[{\bm q, \bm m}]\ ,
\label{f:Phi}
\\
G[{\bm q, \bm m}]&=&\sum_{ab}^{1,n}g(q_{ab})+\ln\det \left({\bm
    q}-{\bm m} \otimes \bm m\right)+\sum_{a=1}^n h(m_a) \label{f:G}\ .
\end{eqnarray}
where
\begin{eqnarray}
g(q) = \sum_{p}\frac{\mu_p}{p} q^p ,&& \qquad \mu_p\equiv
\frac{p\beta^2J_p^2}{2}\, ,
\\
h(m)=\sum_p \frac{\nu_p}{p} m^p, &&\qquad \nu_p\equiv 2 p\beta S_p\, ,
\end{eqnarray}
$\bm q=\{q_{ab}\}$ is
the (symmetric) overlap matrix, $\bm m=\{m_a\}$  the magnetization vector in the replica space and 
$\otimes$ represents the outer product of
vectors. The free energy  $G[\bm q,\bm m]$ must be evaluated at the solution of the saddle
point equations
\bea
\frac{\partial G[\bm m,\bm q]}{\partial q_{ab}}=0\ \ \quad a \neq b\, ,\\
\frac{\partial G[\bm m,\bm q]}{\partial m_{a}}=0.
\eea
When breaking the permutation symmetry among replicas 
an {\em Ansatz} is imposed on the structure of overlap matrix
${\bm q}$. For a generic $R$-steps of replica symmetry breaking \cite{Parisi80} the matrix
${\bm q}$ is divided along the diagonal into successive boxes of decreasing size.
In each box $q_{ab}$ is constant and takes the value, from larger to smaller boxes,
 $0\leq q_0<q_1<\ldots<q_R<q_{R+1}=1$.
The saddle point equation are better written with the help of the two additional functions:
\be
\Lambda(q)=g'(q), \ \qquad \omega(m)=\frac{h'(m)}{2}\, .
\ee
where the "prime" denotes the derivative.

\subsection{RS and 1RSB solutions}
We first consider the case of no replica symmetry breaking $R=0$,
the replica symmetric (RS)  {\em Ansatz}.
The free energy (\ref{f:Phi}) then becomes:
 \begin{equation}
 -2\beta F(\beta) = 1+\ln 2\pi+g(1)-g(q_0)
 +\log (1-q_0)+\frac{q_0-m^2}{1-q_0}+h(m)
 \label{FE_RS}
 \end{equation}
while  for the internal energy we have
 \begin{equation}
 e(\beta)=-T\,[g(1)-g(q_0)]-\frac{T}{2}h(m) \label{RS_energy}.
 \end{equation}
 The parameters $q_0$ and $m$ are obtained from the  self-consistent equations:
\begin{equation}
 \Lambda(q_0) =  \frac{q_0-m^2}{(1-q_0)^2}, \ \qquad \omega(m)  =
 \frac{m}{1-q_0}\,  .
  \label{saddle}
\end{equation}
The particular RS solution $m=q_0=0$ gives the paramagnetic solution.

The RS solution is stable if and only if
\be
\Lambda'(q_0)(1-q_0)^2<1\, .
\ee
This inequality is always satisfied for large enough temperature $T$. However
as $T$ is lowered the RS solution may become unstable and replica symmetry must be broken.
The type of replica symmetry breaking, that is the value of $R$, depends
on the terms appearing in the sum (\ref{Hamiltonion}). If we limit ourself to the case
3+4-FM than only one step of replica symmetry breaking is 
needed \cite{CrisantiLeuzzi07,CriLeu12}, see also the Appendix, thus  $R=1$ leading
to the one-step RSB (1RSB) {\em Ansatz}.
The free energy for 1RSB {\em Ansatz} reads:
\bea
-2\beta F(\beta) & =  &g(1)-g(q_1)+x\,\bigl[g(q_1)-g(q_0)\bigr]+\frac{1}{x}\ln \chi(q_0)
\nonumber\\
 &\phantom{=}&\phantom{==} +\frac{x-1}{x}\ln \chi(q_1) +\frac{q_0-m^2}{\chi(q_0)}\, ,
 \label{f:FEN_1rsb}
\eea
where
\bea
\chi(q_1) & = & 1-q_1\, ,\\
\chi(q_0) & = & 1-q_1+x(q_1-q_0)\, .
\label{f:chi0}
\eea
The parameter
$x$, called replica symmetry breaking parameter, represents the fraction of states 
(replicas)
with overlap  $q_1$. 
Similarly for the 1RSB energy we have
\be
e(\beta)=-T\Bigl[g(1)-g(q_1)+x\,\bigl[g(q_1)-g(q_0)\bigr]\Bigr]-\frac{T}{2}h(m).
\ee
The order parameters $q_1,q_0,m,x$ are the solutions of the  saddle point equations:
\bea
& & \Lambda(q_0)  =  \frac{q_0-m^2}{\chi^2(q_1)}\, ,
\label{f:sp0}
\\
& & \Lambda(q_1)-\Lambda(q_0) =  \frac{q_1-q_0}{\chi(q_1)\chi(q_0)}\, ,
\label{f:sp1}
\\
& & \omega(m)=\frac{m}{\chi(q_1)}\, 
\label{f:spm}
\eea
and the additional requirement of stationarity of $G[q_0,q_1,x,m]$ with respect to variation of $x$
\begin{equation}
g(q_1)-g(q_0)  =
-\frac{1}{x^2}\ln\[[\frac{\chi(q_1)}{\chi(q_0)}\]]+(q_1-q_0)\[[\frac{q_0-m^2}{\chi(q_0)^2}-\frac{1}{x\chi(q_0)}\]].
\label{f:spx}
\end{equation}
The 1RSB solution is stable as long as
\be
\Lambda'(q_1)\chi^2(q_1)<1\, 
\label{stability_q1}
\ee
and
\be
\Lambda'(q_0)\chi^2(q_0)<1
\ee
are satisfied.

In systems with discontinuous 1RSB, besides the thermodynamic phase transition
the replica theory also predicts a dynamic 1RSB (d1RSB) transition.
The d1RSB solution can be obtained in the framework of the
replica calculation using
\eref{f:sp0}-\eref{f:spm}  and a different equation in place of \eref{f:spx}. The latter is the so-called 
marginality condition for dynamic arrest and requires that $q_1$ be marginally stable, that is
\bea
\Lambda'(q_1)=\frac{1}{(1-q_1)^2}
\label{f:dyn_marg}
\eea
that ensures that the characteristic time of the correlation function diverges at, and below, the 
transition. This condition can be derived in different ways:  (i)
requiring that the complexity functional counting excited metastable states is maximal \cite{CrisantiSommers91,CrisantiLeuzzi03,CrisantiLeuzzi06}, or, equivalently, (ii) imposing that the lowest stability eigenvalue of the replica solution tends to zero \cite{CrisantiSommers91,CrisantiLeuzzi06}.
Else, (iii) it can be obtained as a saddle point equation for the RS solution with $q=q_1$ sending the number of replicas $n\to 1$ rather than $n\to 0$ \cite{FranzParisi95,FranzParisi97,Crisanti08} or, eventually, (iv) directly solving the equilibrium dynamics \cite{CrisantiHornerSommers93,CrisantiLeuzzi07b}.

From the Legendre transform of
\eref{f:FEN_1rsb} with respect to $\beta x$, the complexity functional can be computed as 
\bea
2\Sigma(q_1) &=& x^2\frac{\partial \beta F}{\partial x} = -x^2\left[ g(q_1)-g(q_0)
\right]+x^2\Lambda(q_0)(q_1-q_0)
\\
\nonumber
&& -x\frac{q_1-q_0}{\chi(q_0)}-\ln \frac{\chi(q_1)}{\chi(q_0)} \, .
\eea
Using \eref{f:sp1} and the definition of $\chi(q_0)$ \eref{f:chi0}, $x$ can be eliminated
 and $\Sigma(q_1)$
 can be rewritten as
\bea
2\Sigma &=&
-1+(1-q_1)^2{\cal L}_{10}-\ln \left[(1-q_1)^2
{\cal L}_{10}
\right]\\
&&-\left[ \frac{1}{1-q_1}-(1-q_1)
{\cal L}_{10}
\right]^2\left[
g(q_1)-g(q_0)-\Lambda(q_0)(q_1-q_0)
\right]\, ,
\\
{\cal L}_{10}&\equiv  &\frac{\Lambda(q_1)-\Lambda(q_0)}{q_1-q_0} \, ,
\nonumber
\eea
where $q_0$ is eliminated in favor of $q_1$ using
\bea
\Lambda(q_0)=(q_0-m^2)(1-q_1)^2
{\cal L}_{10} \, .
\eea

\subsection{The 3+4-FM model: phase diagram and complexity}
\label{sec:LucaProposal}
\begin{figure}[t!]
\center
\includegraphics[width=.95\textwidth]{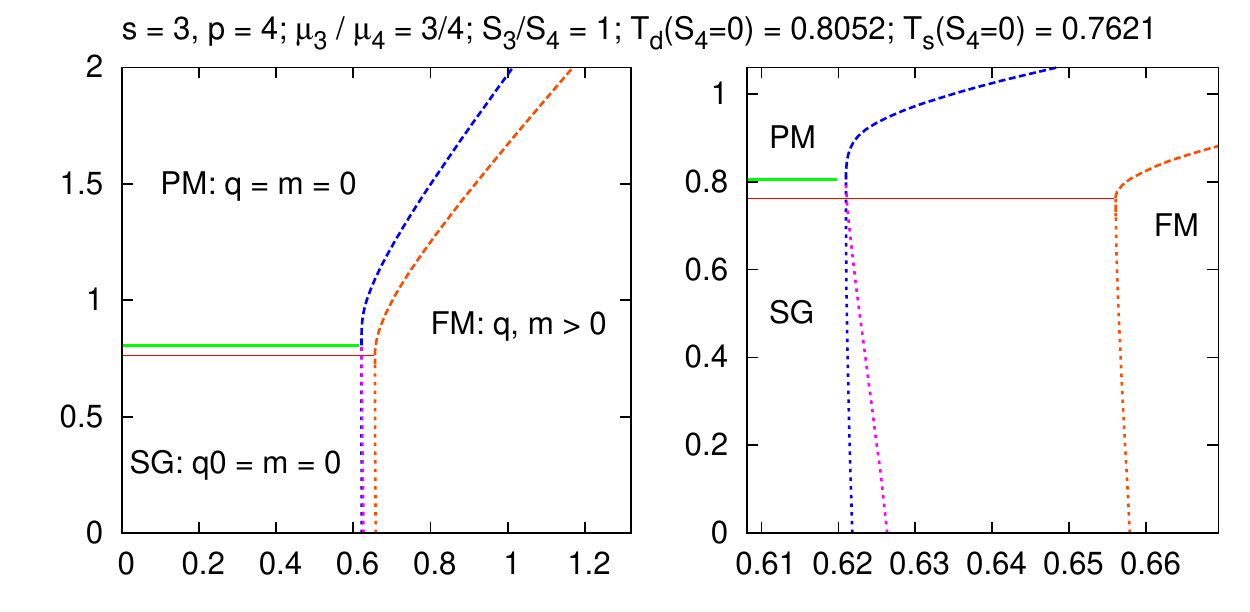}
\caption{\small{Phase diagram of the 3+4-FM spherical spin glass model
          with $J_3=J_4$ and $S_3=S_4$. 
          The lines are:
           Red full:  static PM/SG  transition;
           green full: dynamic  PM/SG transition; 
           red dashed:  PM/FM transition;
           red dotted: SG/FM  transition;
           blue dashed: FM spinodal;
           blue dotted: FM$_{\rm 1RSB}$ spinodal;
           magenta dotted:  FM$_{\rm 1RSB}$/FM transition.
           The latter is hardly seen on the left panel.
           Right: Detail of  the spinodal lines. 
             }
           }
\label{fig:3+4_PhDi}
\end{figure}

In this section we shortly summarize some features of the spherical 3+4-FM  spin glass model that will 
be employed to test the states following procedure in the rest of the paper.
In the following we choose, without loss of generality, 
$S_3=S_4$ and $J_3=J_4$.
The phase diagram is displayed in Fig.~\ref{fig:3+4_PhDi} and consists of three equilibrium phases  
separated by  red lines.
At high temperature $T$  and not too large $S_4$  the phase is paramagnetic (PM) with $q=m=0$, 
while at low $T$ and sufficiently small  $S_4$
we have a spin glass (SG) phase of 1RSB type with $q_0=m=0$.
For large $S_4$ the phase becomes ferromagnetic (FM) of RS type with both $m$ and $q$  positive. 
The transition to the FM phase is of first order, and the blue lines depict the  
spinodal lines of the FM and FM$_{\rm 1RSB}$ --ferromagnetic of 1RSB type-- metastable phases. 
Note that the FM$_{\rm 1RSB}$ metastable phase goes
continuously over a FM metastable phase, magenta line, 
before reaching the ferromagnetic transition.
The SG transition is a discontinuous random first order transition, occurring at the Kauzmann 
temperature $T_K$, also called static temperature $T_s$,  horizontal red line.
We shall use both notations depending on the context.
Finally the horizontal green line gives the dynamical transition that occurs at the 
higher dynamical transition temperature $T_d$, also called the mode coupling critical temperature.

\begin{figure}
\center
\includegraphics[width=.49\textwidth]{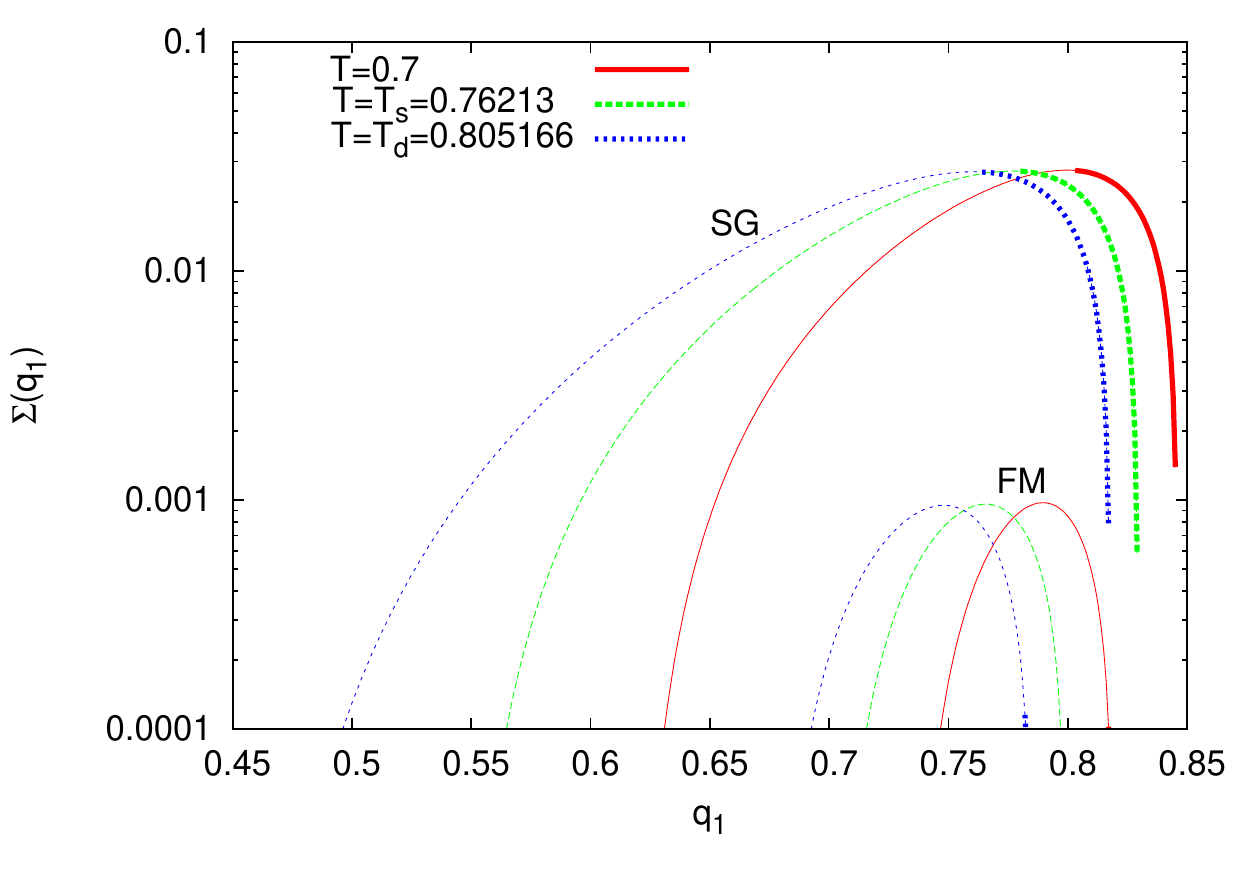}
\includegraphics[width=.49\textwidth]{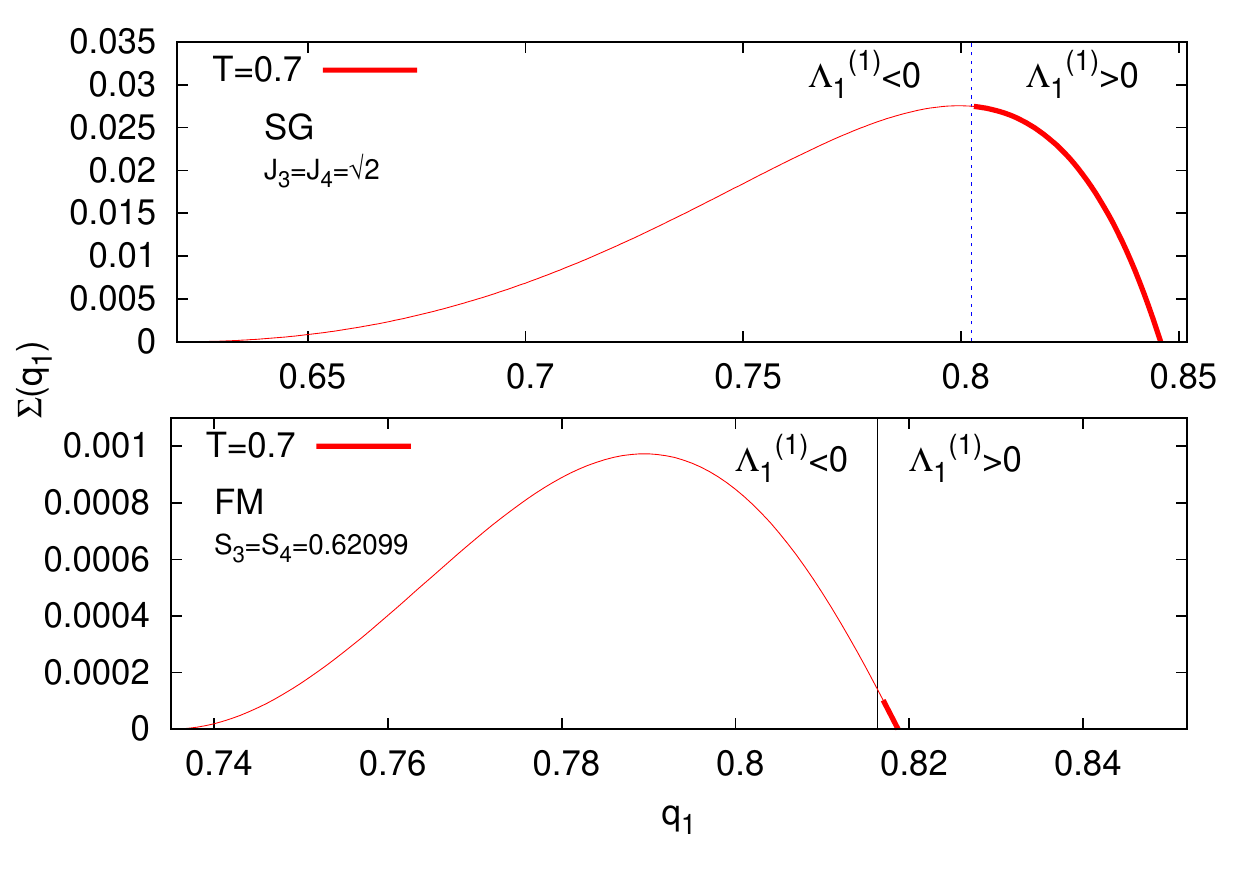}
\caption{\small{Left: Complexity curves versus $q_1$ for the spherical 3+4-FM spin glass model
                           with $J_3=J_4$ and $S_3=S_4=0.62099$. 
                           Left: $\Sigma(q_1)$ for  $T=0.7$, $T=T_k=0.76213$ and $T=T_d=0.80517$ 
                           in the SG phase and in the FM$_{\rm 1RSB}$ phase.
                           Plot is in $y$-log scale to allow for comparison of complexity curves in both phases.
                           Thick lines, hardly seen for the FM$_{\rm 1RSB}$ phase, represent stable branches, 
                            thin lines unstable. 
                           Right: $\Sigma(q_1)$ at $T=0.7$ for the SG  states (top) and for the FM$_{\rm 1RSB}$ states
                            (bottom). 
                            Vertical lines denote the instability point.
                            }}
\label{fig:complex_FM_SG}
\end{figure}

When the ferromagnetic FM$_{\rm 1RSB}$ phase exists, its complexity,
i.e., the logarithm of the number of metastable states,  is extensive.
We can then compare it with the complexity of the SG phase.
The simplicity of the spherical model allows to plot them as function of $q_1$ only. 
This is done in \Fref{fig:complex_FM_SG} where the behavior of $\Sigma(q_1)$  versus $q_1$ 
is shown for three temperatures:  $T<T_K$, $T=T_K$ and $T=T_d$. 
Not all values of $q_1$ correspond to physical solutions, i.e., satisfy Eq. (\ref{stability_q1}): only the
thick lines refer to the interval in which $\Sigma(q_1)$ counts physical solutions.
The thin lines counts solutions that are unstable in the replica space, and hence unphysical. 
One can notice that the order of magnitude of the complexity of the FM$_{\rm 1RSB}$ 
is two orders of magnitude smaller than that of the spin-glass phase.

\section{ Equations for evolution of states for $T_e \geq T_K$}
\label{following}

In this section, we give the equations for adiabatic evolution of
equilibrium states at $T_K \leq T_e \leq T_d$ in temperature $T_a$
following the planting procedure as introduced in
Ref.~\cite{KrzakalaZdeborova09}.  We recall that the temperature $T_K$
(or $T_s$) is the so-called Kauzmann temperature at which a true
static thermodynamic glass transition takes place, at least in
mean-field systems, between a paramagnet and a thermodynamically
stable equilibrium glass (the {\em ideal} glass).  For simplicity, we
only consider models with $S_p=0$.

For $\beta_e \geq \beta_K$ (that is, $T_e \leq T_K$), in order to
select an equilibrium configuration, we, first, generate a
configuration randomly and then construct all of the interaction
constants $J_{i_1,i_2,\ldots,i_r}$ so that the energy of the planted
configuration equals $e(\beta_e)$, cf. Eq.  \eref{RS_energy}.  The
important point is that, since above the Kauzmann temperature the RS
quenched free energy is exactly the annealed free energy, the random
planting ensemble is equivalent with the usual random ensemble,
i.e. we can do a quiet planting
\cite{KrzakalaZdeborova09,KrzakalaZdeborova11}. This is a very non
trivial property: as long as $T>T_K$ when we generate a random
configuration and a random set of interactions, so that this
configuration is an equilibrium one, we have simply generated an
equilibrated configuration of a typical realization of the usual
random ensemble.

It seems difficult to plant an
equilibrium configuration in the spherical spin models due to the
fact that $\{\sigma_i\}_{i=1}^N$ are real variables which can range from
$-\infty$ to $\infty$.  However, the RS free energy of the spherical
model with $S_p=0$ equals that of Ising model
\cite{ZdeborovaKrzakala10}, which means that the Gibbs measure of
spherical model is dominated by the Ising configurations $\{-1,1\}^N$.
Consequently, for the spherical model, we can follow the same planting
procedure as introduced in the Ising model in order to derive the
equations of the evolution of states. We refer the reader to
\cite{KrzakalaZdeborova11} for details, but the point is that for
such models where RS quenched free energy is the annealed free energy,
one can map the following state procedure to a usual static
computation in a model with a ferromagnetic bias. We shall briefly
recall how this can be done:

First, we generate a random configuration $\bf{\sigma}$ with $\sigma_i
\in \{-1,1\}$. We then generate the (planted ensemble) interactions
according to
\be P(J|{\bf{\sigma}}) \propto e^{-\beta {\cal{H}}_J ({\bf \sigma})}
\prod_{ij} {\cal {N}} \left( J_{ij}, 0 , \frac{J_p^2 p!}{2N^{p-1}} \right), \ee
where ${\cal N}(x,\bar x,\Delta)$ is the normal distribution of mean $\bar x$
and variance $\Delta$. In doing so, we have generated a typical
configuration of the planted ensemble, which is the same as the
quenched one as long as $T>T_K$.  In the annealed spherical model with
$S_p=0$, the interaction variance scales as
$\langle{J^{(p)}_{i_1\ldots i_p}}^2\rangle=J_p^2 p!/(2N^{p-1})$, {
  implying an extensive, $O(N)$, energy and}, thus, from the central
limit theorem, we have the typical value $J^{(p)}\sim J_p
\sqrt{p!/(2N^{p-1})}$. 

Call a
  $p$-interaction is satisfied (unsatisfied) if the energy contribution $J^{(p)}_{i_1,i_2,\cdots
  i_p} \sigma_{i_1}\sigma_{i_2}\cdots\sigma_{i_p}$ is negative (positive).
Then, at inverse temperature $\beta_e$, we have the following
expression for the fraction of unsatisfied interactions:
\begin{equation}
\epsilon^{(p)}=\frac{1}{2}+\frac{e(\beta_e)}{J_p}\sqrt{\frac{p!}{2N^{p-1}}}
\end{equation}
where $e(\beta_e)$ is defined in \eref{RS_energy}. As we only consider $\beta_e
\leq \beta_K$, in the
model with $S_p=0$, it leads to $q_0=m=0$ and therefore,
\begin{equation}
\epsilon^{(p)}=\frac{1}{2}-\frac{\beta_e J_p}{2}\sqrt{\frac{p!}{2N^{p-1}}}.
\end{equation}

We now use the fact that this configuration can be transformed to a
uniform one (all $\sigma_i=1$) due to the Gauge invariance, i.e., for
any spin $i$, the transformation $\sigma_i \to -\sigma_i$, $J_a \to
-J_a$ (for all interactions $a$ involving spin $i$) will keep the
Hamiltonian \eref{Hamiltonion} invariant. Then, for all $p$, choose
the signs of interactions in the $p$-interactions:
$J_{i_1,i_2,\ldots,i_p}^{(p)}$, such that fraction $\epsilon^{(p)}$ of
them is unsatisfied.

After this transformation is performed, we are left with the problem of finding where the
uniform "all up" configuration is an equilibrium one, and where
the distribution of interactions is given by 
\be P^{\mathrm{eff}} (J) \propto \prod_{ij} {\cal {N}} \left(J_{ij},
\frac{S_p^{\mathrm{eff}} p}{N^{p-1}},  \frac{J_p^2 p!}{2N^{p-1}} \right), \ee
with
\begin{equation}
S_p^{\mathrm{eff}}=\frac{\beta_eJ_p^2}{2}. \label{Nishi}
\end{equation}
In fact, \eref{Nishi} is exactly the Nishimori line condition for
spherical $p$-spin spin glass model \cite{ZdeborovaKrzakala10}. Therefore, for
model with $S _p=0$ and for $T_e>T_K$, all of the thermodynamic
properties of the states can be obtained from the phase diagram of the
corresponding ferromagnetically biased model. In particular, since the
planted configuration is an equilibrium one at $T_e$, we can rewrite
the complexity function (i.e., the logarithm of the number of
equilibrium thermodynamic states) at $T_e$ as following:
\begin{equation}
\Sigma(\beta_e)=-\beta_eF(\{S_p=0\} )+\beta_eF(\{S_p^{\mathrm{eff }}\})\label{complexity}
\end{equation}
from which we can determine the dynamical temperature $T_d$, i.e., the
highest temperature at which Eq.  \eref{complexity} becomes positive, and the
Kauzmann temperature $T_K$,  at which  Eq.~\eref{complexity} goes to 
zero. The reader interested in this correspondance between the
following state procedure and the Nishimori line is referred to
\cite{KrzakalaZdeborova09,ZdeborovaKrzakala10,KrzakalaZdeborova11} for
more details.

\section{Results and discussion}
\label{results}
In this section we present the results of adiabatic state following for the spherical 3+4-FM spin glass
model with $J_3=J_4=1$, and $S_3=S_4=0$, we discuss their physical meaning and compare with
the existing results.

\subsection{Mapping to the phase diagram of the ferromagnetically
  biased model}
\begin{figure}[t!]
\begin{center}
\includegraphics[width=0.8\linewidth]{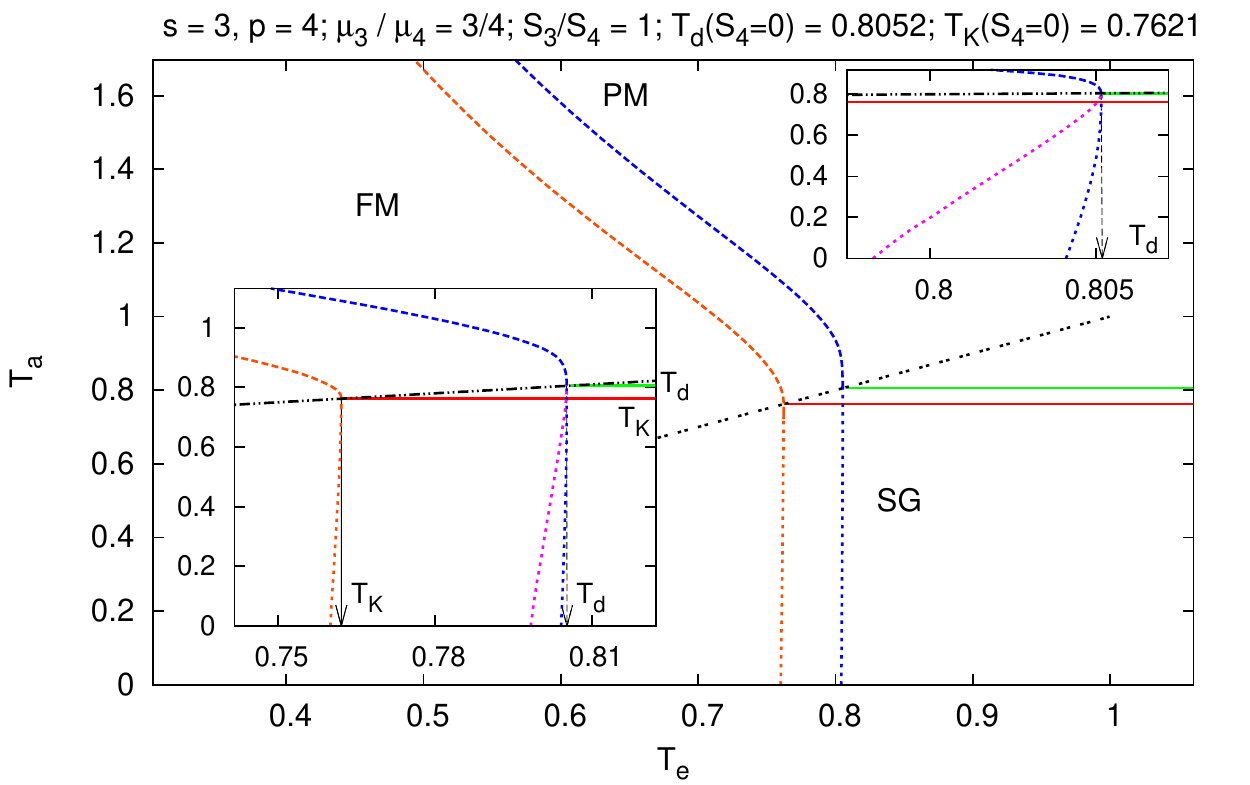}
\caption{\small{Phase diagram of the spherical 3+4-FM spin glass model as a function of
                temperature $T_a$, and $T_e$ $(2T_e=1/S^{\rm eff}_4)$.
                Three thermodynamic phases are present separated by red lines:
                 PM, SG  an FM.
                 The blue line is the spinodal line, while 
                 the green line is the dynamic PM/SG  transition. 
                 For more details on the lines see Fig.~\protect{\ref{fig:3+4_PhDi}}.
                 Note that both the FM phase transition and the spinodal line 
                 are reentrant, as emphasized in the insets. 
                 The metastable state that appears at the spinodal line is either FM or FM$_{\rm 1RSB}$.
                 The magenta line denotes the transition between these two phases.
	} }
\label{phase}
\end{center}
\end{figure}

By exploiting  the equivalence between the evolution of states
that were the equilibrium ones above $T_K$ and the thermodynamic properties of the model
on the Nishimori line, we can acquire all the information about how
states evolve with temperature from the phase diagram of the spherical 3+4-FM spin glass model
discussed in Sec.~\ref{sec:LucaProposal} and plotted in \Fref{fig:3+4_PhDi}. 
Due to the Nishimori condition we will now have $S^{\rm eff}_3=S^{\rm eff}_4=\beta_e/2$. 
We show the phase diagram in the $T_e$, $T_a$ space
in \Fref{phase}, where   the
Nishimori line, given by $T_e=T_a$, is plotted as a black line:
it intersects the spinodal transition line at the
dynamical temperature $T_d$ and it exactly crosses the intersection of three phase transitions
lines at the Kauzmann temperature, $T_a=T_e=T_K$.
In the inset of 	\Fref{phase} we emphasize the re-entrant nature of the FM
phase, comparing the borderline of  the FM phase (red dotted) and the FM
spinodal line (blue dotted) with vertical arrows, respectively at $T_e=T_K$ and $T_e=T_d$.
The magenta line separates the FM and FM$_{\rm 1RSB}$ (metastable) phases.

To interpret this phase diagram for the adiabatic evolution of Gibbs
state we consider states that were at equilibrium at temperature $T_e$, $T_K \leq T_e
\leq T_d$. The Nishimori line $T_a=T_e$ is hence the equilibrium line
for $T_e>T_K$. As $T_a$ increases the states encounter the
ferromagnetic spinodal (blue dash-dotted line) and ``melt'' into the
paramagnet. As $T_a$ decreases there are three possible cases
depending on the value of $T_e$ between $T_K$ and $T_d$, c.f.r. \Fref{phase}.
Starting from  $T_e=T_K$, and increasing $T_e$, we have:

\begin{itemize}
    \item For $T_e>T_K$, the
      state can be followed lowering $T_a$ using the RS {\em Ansatz} down to
      zero temperature.
    \item As $T_e$ becomes larger than the value where the FM/FM$_{\rm 1RSB}$ transition
              line touches the $T_a=0$ 
              axis, see  Fig.~\ref{phase}, then the state can be followed
             using the RS {\em Ansatz} only down to  a bifurcation temperature  $T_a=T_{\rm1RSB}$, 
              where the crossing with the FM/FM$_{\rm 1RSB}$ transition line occurs. Here the state 
            splits into exponentially many sub-states with a 1RSB structure. 
            This ensemble of states can be followed down to zero
      temperature using the 1RSB scheme.
    \item Finally when $T_e$  becomes larger than the value where the FM spinodal line touches
    	     the $T_a=0$ axis, the state can be followed as before with RS {\em Ansatz} down 
	     to $T_{\rm 1RSB}$, where the crossing with the FM/FM$_{\rm 1RSB}$ transition occurs, 
	     and below this point with the 1RSB {\em Ansatz}. However at difference with 
	     the previous case we cannot reach $T_a=0$, but we have to stop at the temperature
	     $T_a = T_{\rm spFM}$ where we cross the reentrant FM spinodal line.
	     Below this temperature ferromagnetic solutions of any type do not exist anymore.
	     Within the state following interpretation of the phase diagram this means that
               the state completely disappears, and hence no solution exists
               correlated with the state in equilibrium at $T_e$. 
               The vanishing of states will be further discussed in section  \ref{KZ_instability}.
\end{itemize}

\subsection{Evolution of states}

\begin{figure}
\begin{center}
\includegraphics[width=.8\linewidth]{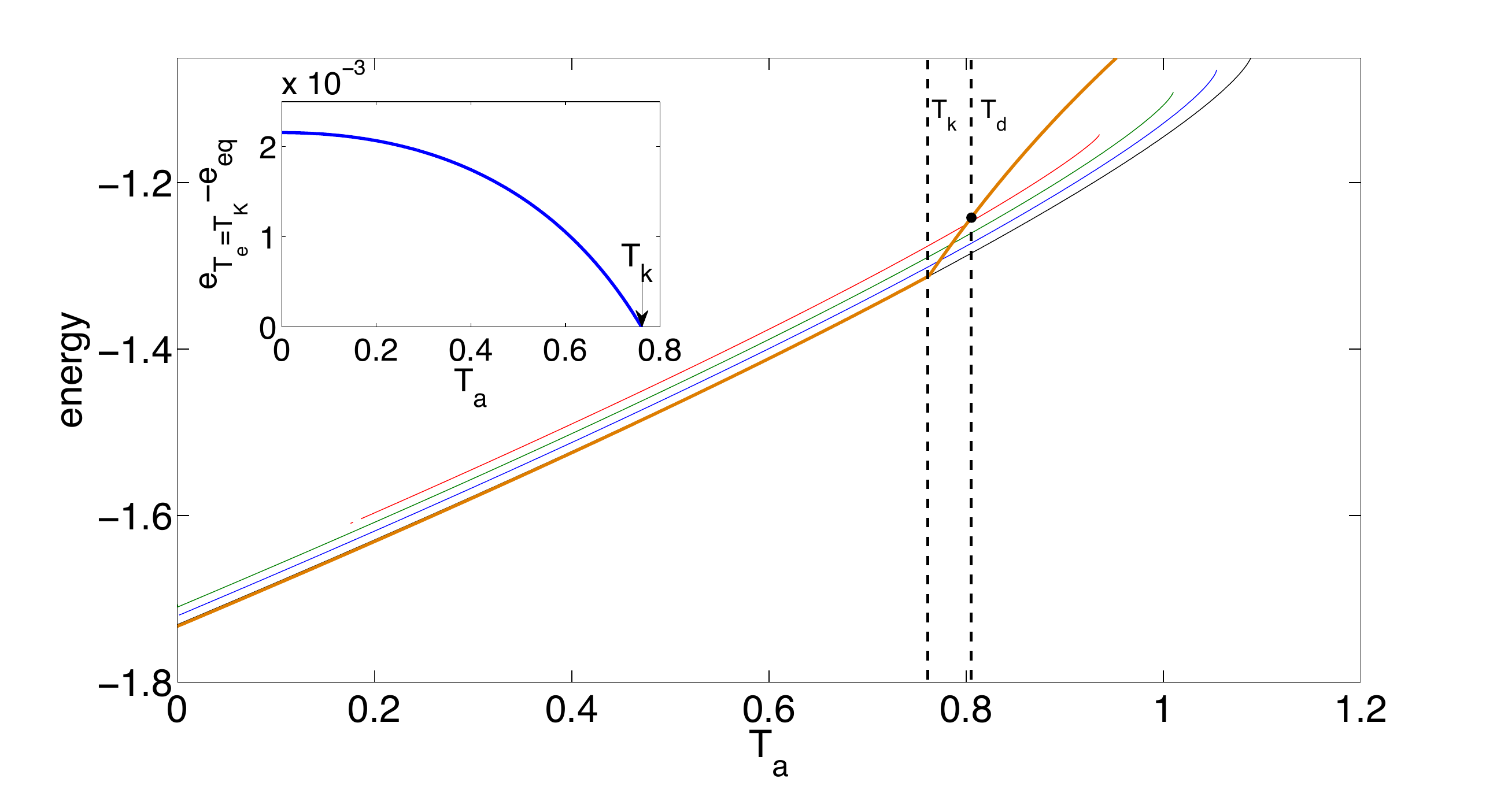}\\
\includegraphics[width=.8\linewidth]{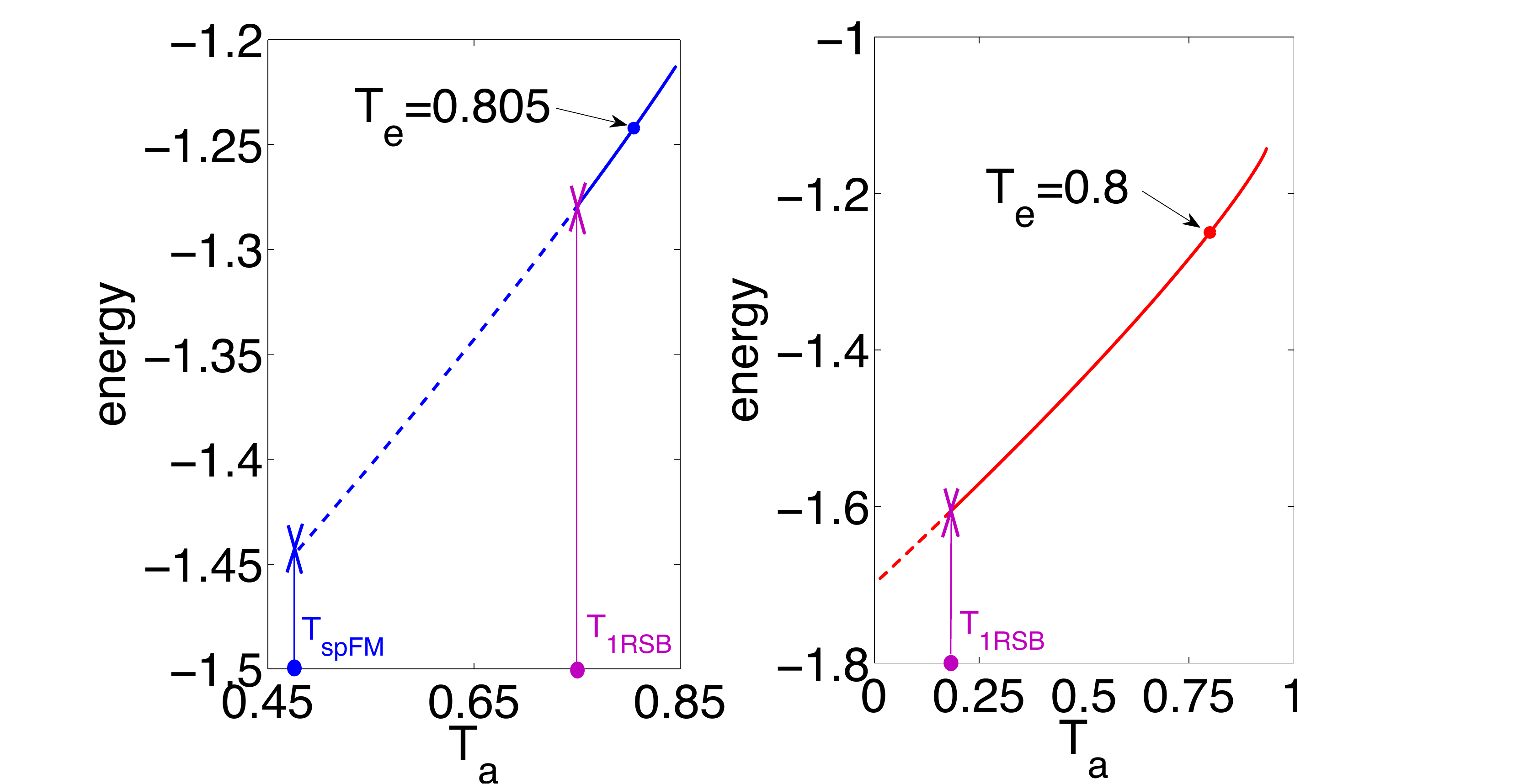}
\caption{\small Evolution of states in spherical 3+4-FM spin glass model
                 for $T_K\leq T_e\leq T_d$. 
                 Top: The RS result; the thick yellow curve is the equilibrium energy $e(T)$ 
                 in the model with $S_3=S_4=0$. Other curves are the evolution of energy of 
                 states with $T_e=0.805166 (=T_d), 0.8,0.785,0.773,0.76213 (=T_K)$, respectively.
                 Notice that the equilibrium state at $T_e=T_d$ disappears as soon at $T_a < T_e$. 
                  Inset: Difference between the equilibrium energy of model and the energy of states 
                  of equilibrium at $T_e=T_K\approx 0.76213$.  
                  Bottom: The result for $T_e= 0.805$ (left) and $T_e=0.8$ (right). 
                  The RS part is drawn in solid line and the 1RSB result in
                  dashed line. For $T_e=0.805$ (left) the state following ends at $T_{\rm spFM}=0.4794$
                  where the reentrance of the  FM spinodal line is reached.
  }
  \label{1RSB_mixed}
\end{center}
\end{figure}

\Fref{1RSB_mixed} shows how the energy of states evolves in
temperature, compared with the equilibrium energy, thick yellow line in figure, 
from the 1RSB computation \cite{CrisantiLeuzzi06}. The top part of
the figure shows the result of state evolution computed within the
RS Ansatz from Eqs.~\eref{saddle}, with $S_p$ given by $S_p^{\mathrm{eff}}$
defined in Eq.~\eref{Nishi}. Upon warming, the energy of the state grows up
to a spinodal temperature beyond which the only solution is 
$(q_0,m)=(0,0)$. As $T_e$ approaches  $T_d$ from below, the energy
of the spinodal point decreases gradually and goes to $e(T_d)$ at $T_e
\to T_d$. Note that this is not the case for the  pure spherical
$p$-spin spin glass model, where one retains only on term in the sum (\ref{Hamiltonion}), 
which is kind of pathological in this aspect. The
states at equilibrium at $T_d$ do not exist at any temperature higher
than $T_d$. Upon cooling, the energy decreases. However, for $T_e$
close enough to $T_d$ the only RS solution of Eqs.~\eref{saddle} at low
$T_a$ is the paramagnetic one. 

The energy of states at equilibrium at
$T_e=T_K$, followed at $T_a<T_K$, is different from the equilibrium
energy and their difference is shown in the inset of \Fref{1RSB_mixed}
(a). This indicates that these states fall out of equilibrium upon
cooling, in contrast with the pure  spherical $p$-spin spin glass
models  where the  equilibrium state at $T_e=T_K$ is always at
equilibrium for $T_a<T_K$. Thus, level crossing and temperature chaos
do appear for $T_e<T_K$ in the spherical 3+4-FM spin glass model.

The lower part of \Fref{1RSB_mixed} shows the evolution of states
computed with the 1RSB Ansatz, which is stable in this
case~\cite{CrisantiLeuzzi06}. The 1RSB solution,
with $(m,q_0)\neq(0,0)$ and $q_1\neq q_0$, corrects the RS result below
the state splitting temperature $T_{\rm 1RSB}$. As illustrated on the left, for $T_e$
close enough to $T_d$ we encounter the reentrant spinodal line, below
which  no solution of the 1RSB equations is found with $m>0$, and
 the state following ends.

 \begin{figure}[h!]
\begin{center}
\includegraphics[width=0.8\linewidth]{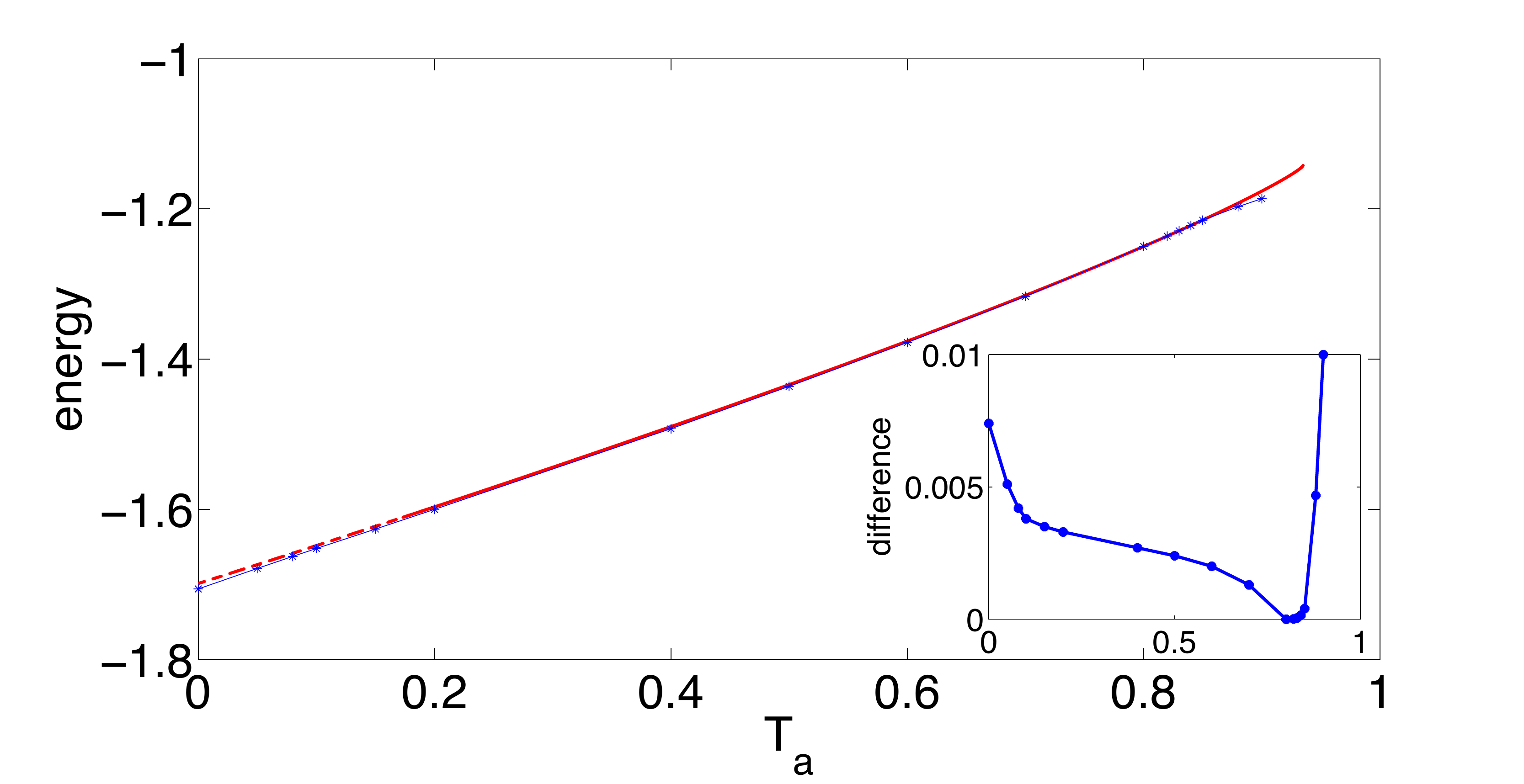}
\caption{Comparison of the results derived from following states method
                (red line) and that from iso-complexity approach \protect{\cite{MontanariRicci04}} 
                (blue line with stars) where $T_e=0.8$. 
                 Inset: The difference between the two energies , 
                 $e_{\mathrm{exact}}-e_{\mathrm{iso}}$,  as a  function of $T_a$. 
                 }
\label{iso_complexity}
\end{center}
\end{figure}

Finally, let us compare our results for adiabatic state following with
an ``iso-complexity'' approximation proposed in Ref.~\cite{MontanariRicci04}
 to study  the cooling procedure in one state.
In the iso-complexity approximation we count the logarithm of the number
of the states at $T_a$ vs. energy and we choose the energy such
that this number equals the equilibrium complexity at $T_e$. The true
state following energy cannot be lower than the iso-complexity energy prediction,
because, otherwise, there would not be enough states at $T_a$ at such
lower energy, but it can be higher, since states that were not the
equilibrium ones at $T_e$ can contribute to the complexity. We show in \Fref{iso_complexity} the energy
of the state at equilibrium at $T_e=0.8$ at different $T_a$ computed
using our exact method and the iso-complexity approximation. As
we expect, the iso-complexity provides a lower bound to the true energy of the state (see the
difference in the inset).   This result is different from the case of
the pure spherical  $p$-spin spin glass model,  in which the two methods give the same results because states do not cross { or disappear.}

 \begin{figure}[t!]
 \centering
 \subfloat[]{\includegraphics[width=3in]{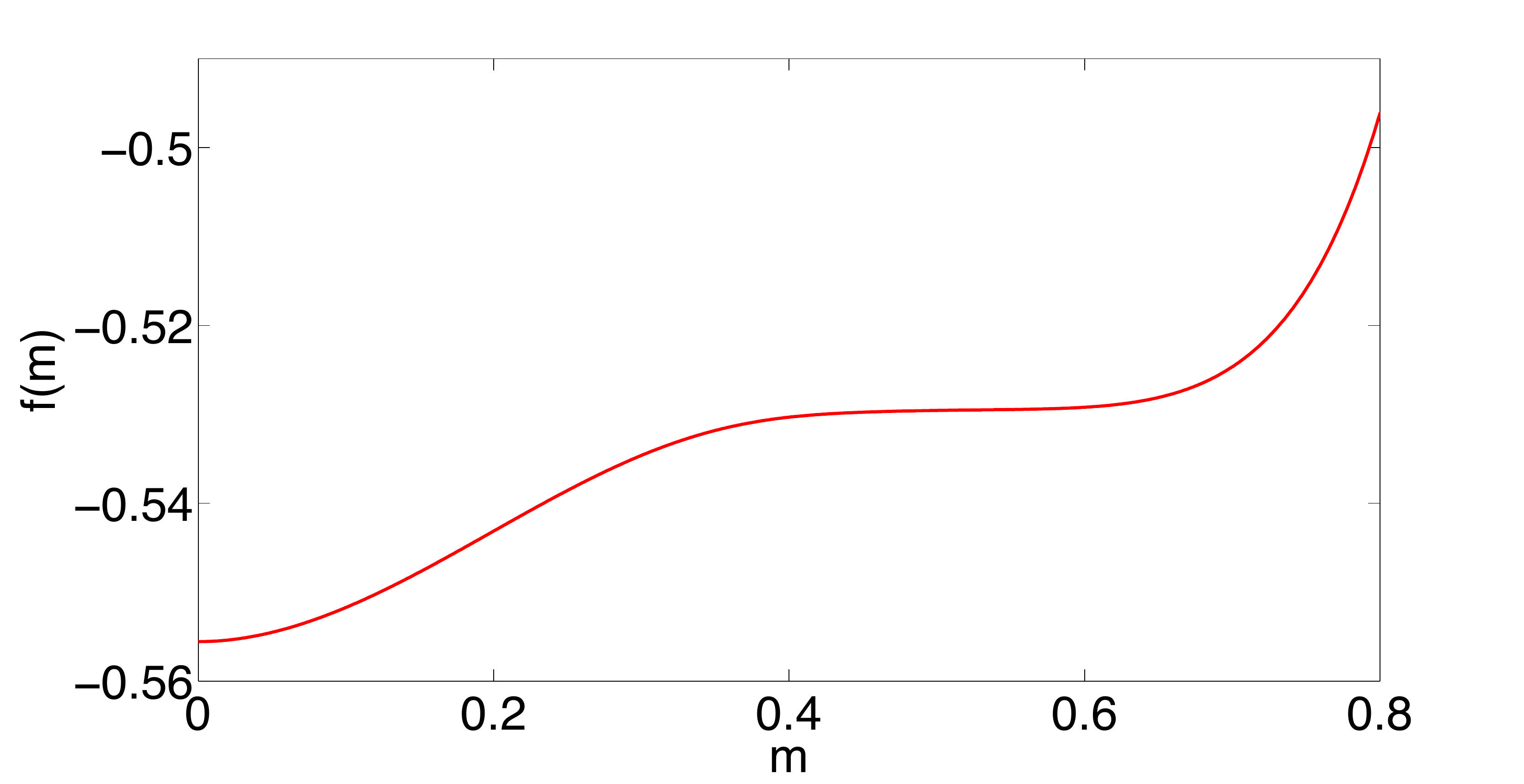}}
 \subfloat[]{\includegraphics[width=3in]{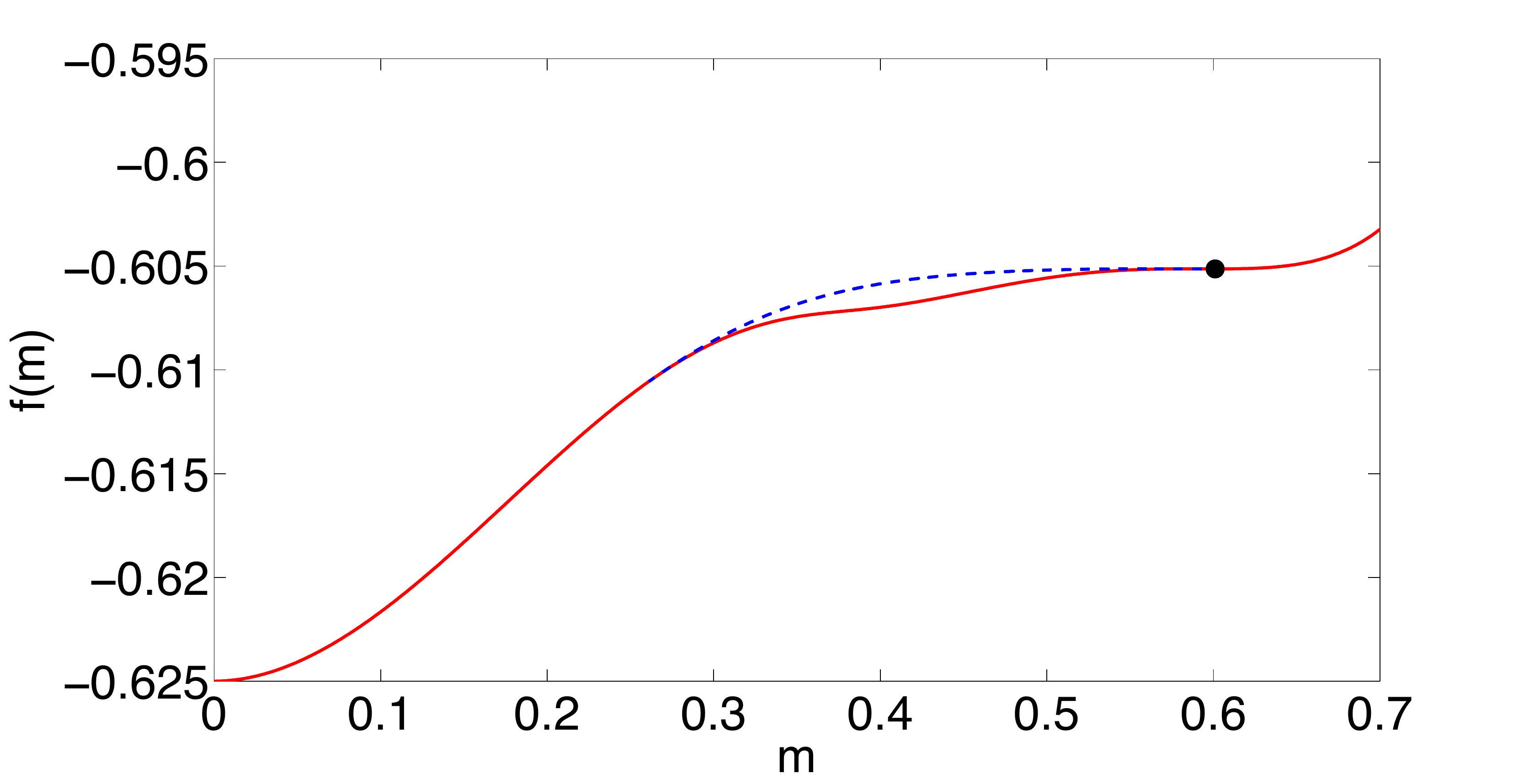}}

  \subfloat[]{\includegraphics[width=3in]{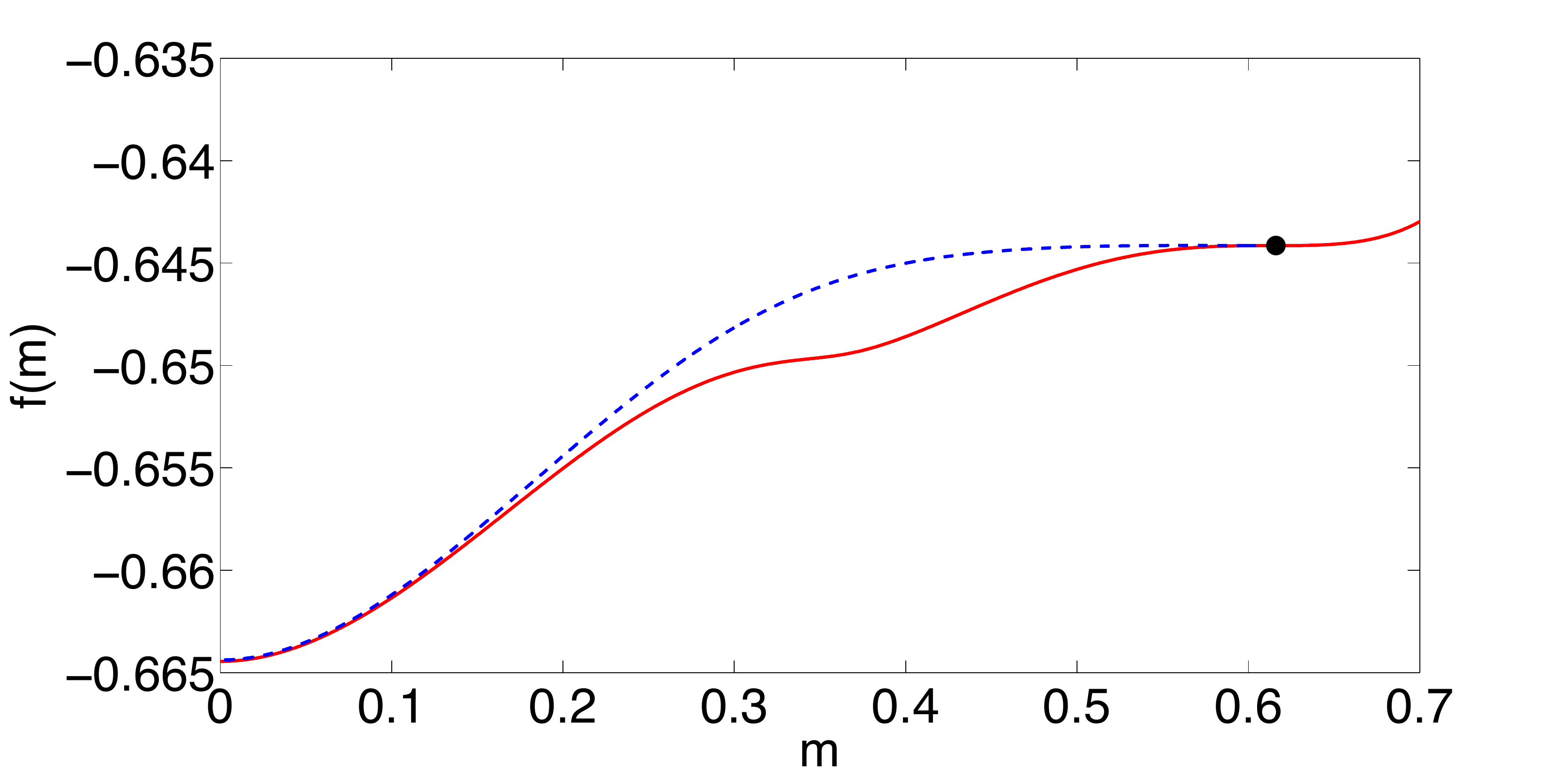}}
   \subfloat[]{ \includegraphics[width=3in]{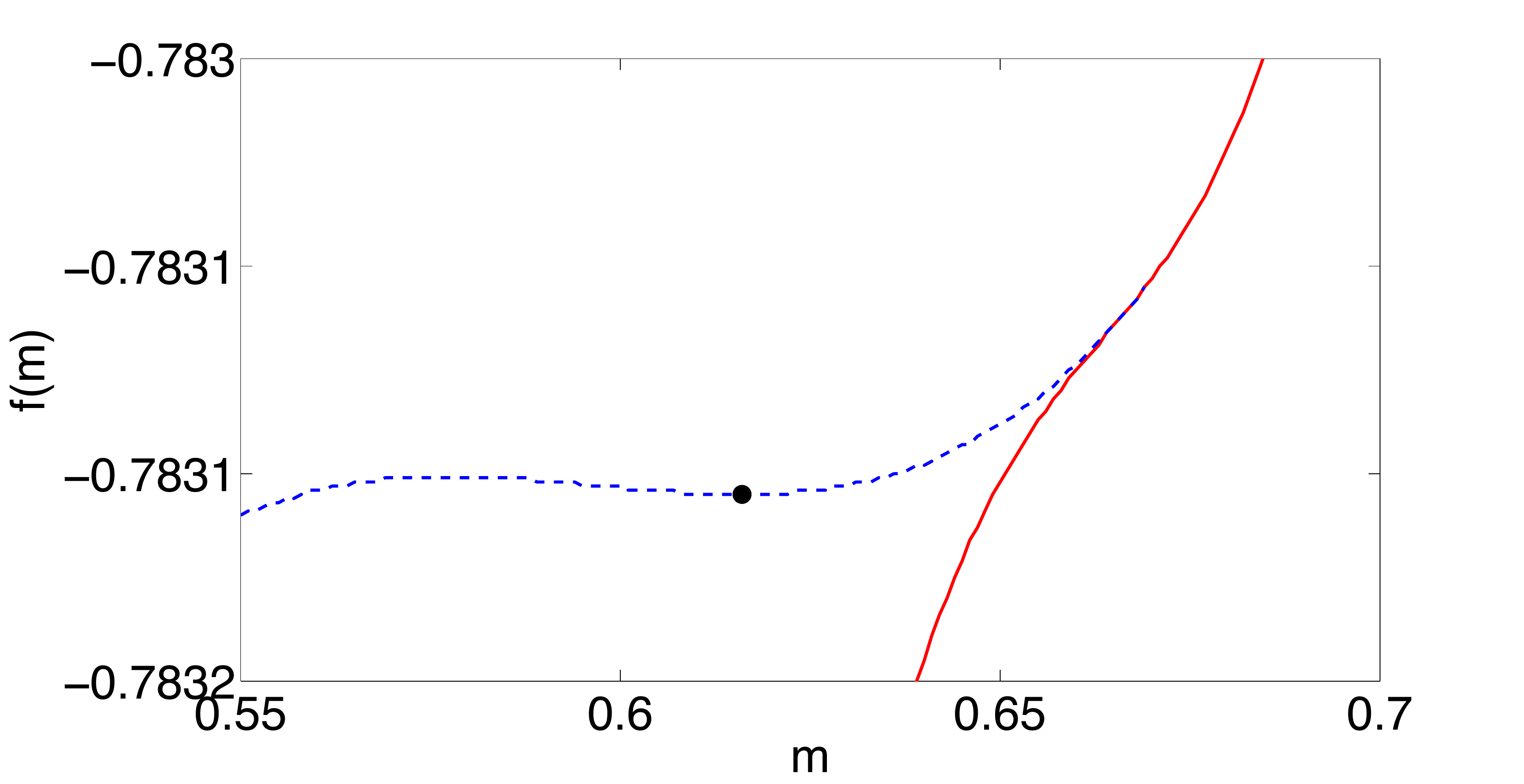}}

  \subfloat[]{\includegraphics[width=3in]{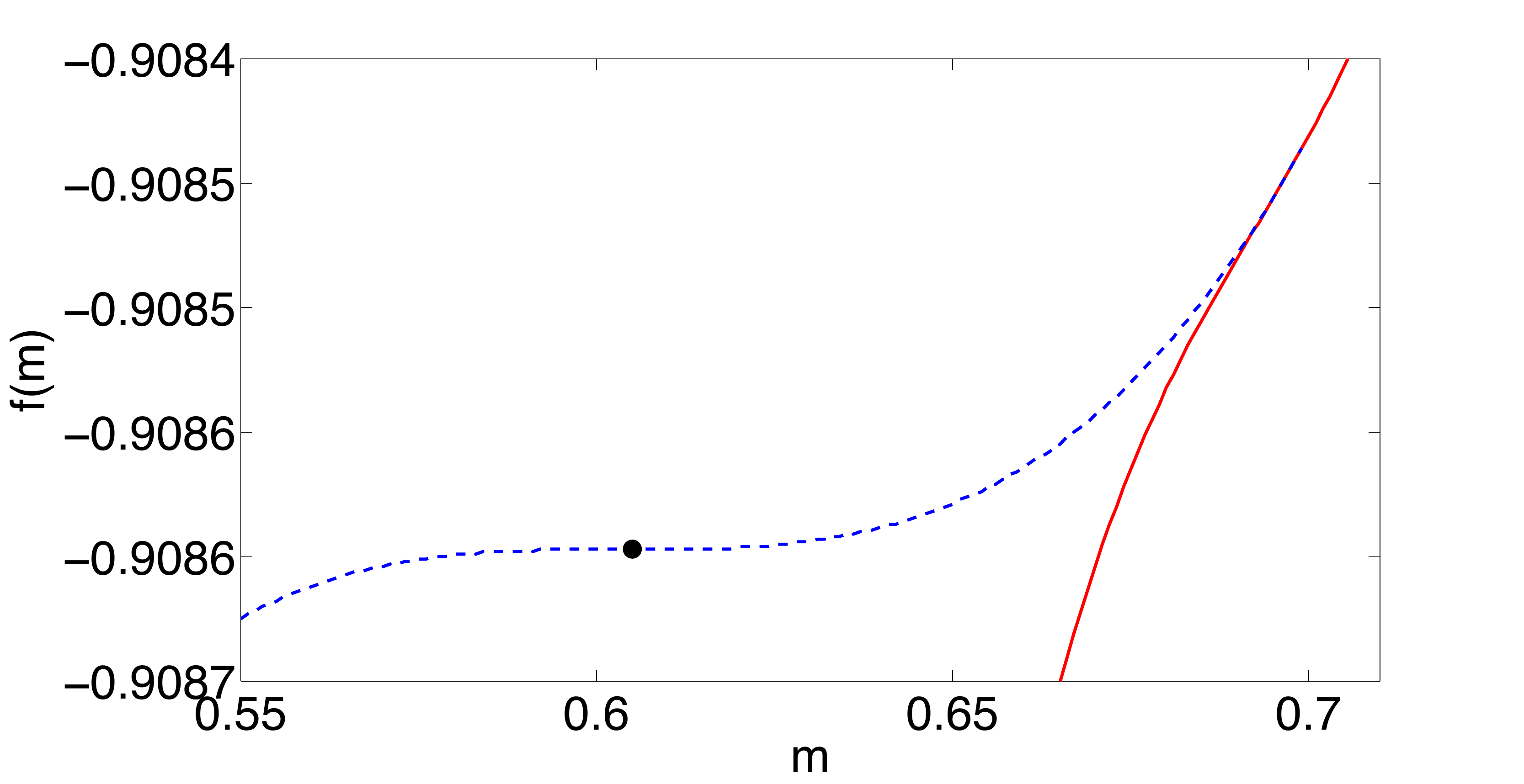}}
 \subfloat[]{\includegraphics[width=3in]{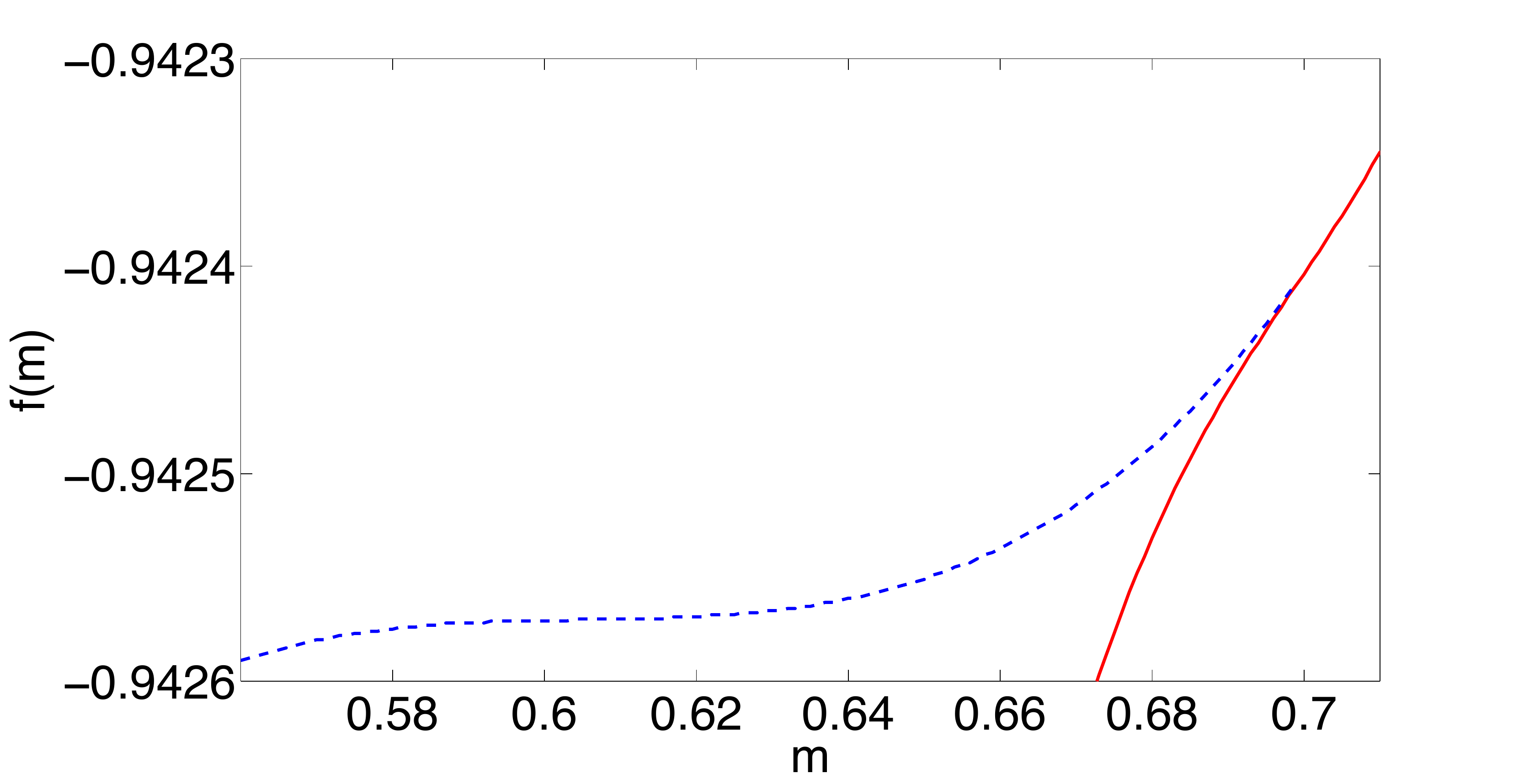}}
\caption{Free energy of a state at $T_a$ at a fixed overlap with a
  configuration at equilibrium at $T_e$. Here, $T_e=0.805$ and $T_a=0.9$ (a),
  $0.805$ (b), $T_{1RSB}=0.7525$ (c), $0.6$ (d), $T_{\rm spFM}=0.4794$ (e), $0.45$ (f),
  respectively. The solid red line represents the RS free energy and
  the dashed line the 1RSB free energy. The marks indicate the minimum
  of the free energy.}
\label{potential}
\end{figure}

\subsection{Relationship with Franz-Parisi potential}
The results presented in the previous sections
are  equivalent to the results of Ref.~\cite{BarratFranz97}. The Franz-Parisi potential measures the free
energy at temperature $T_a$ of the system at fixed overlap with a solution sampled from
equilibrium at temperature $T_e$. Under our planting mechanism, magnetization is exactly the 
overlap between the configuration
in the state of equilibrium at $T_e$ and the typical configuration
in the same state at $T_a$; therefore, the free energy of the model
with a ferromagnetic bias at a fixed magnetization can be directly translated
into the Franz-Parisi potential.

\begin{figure}[!t]
 \begin{center}
 \includegraphics[width=0.49\linewidth]{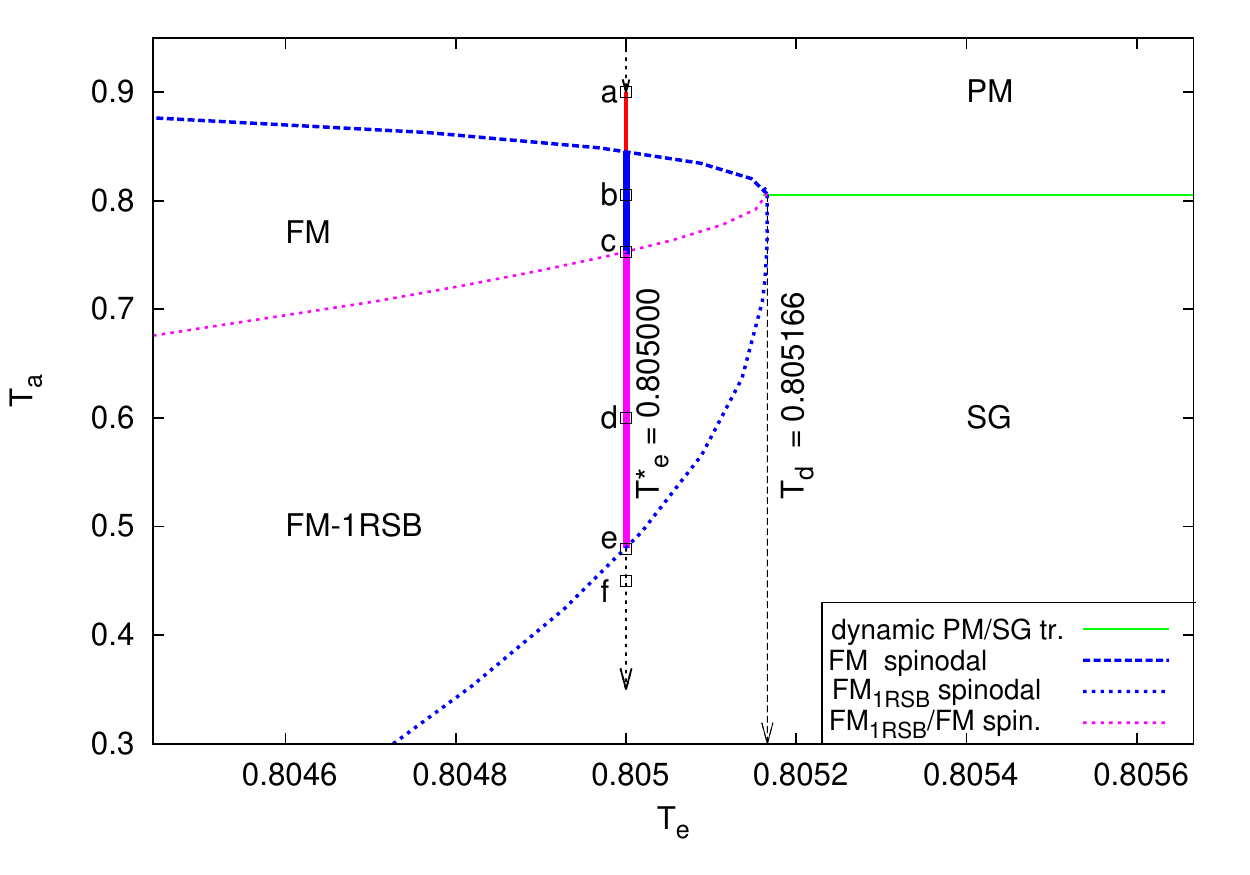}
 \includegraphics[width=0.445\linewidth]{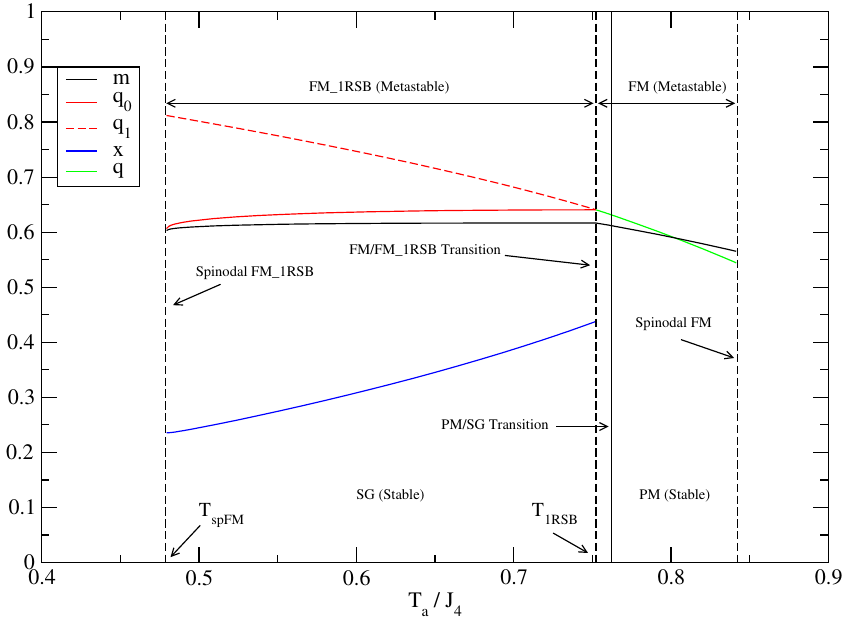}
\caption{\small{Left: Zoom into the $T_a$, $T_e$ phase diagram. 
                The cooling path at $T_e=0.805$ is denoted by the full-line. The temperatures $T_a$
                shown in  Fig.~\protect{\ref{potential}} are indicated by squares. 
                Right:  The magnetization $m$ (black solid line) and the parameters $q$ (RS) and 
                 $q_0$, $q_1$ and $x$ (1RSB) as function of $T_a$ along the cooling path
                 $T_e=0.805$.
                 The magnetization $m$ coincides with the local minimum $m^{\star}$ of the free
                 energy shown in Fig.~\protect{\ref{potential}}.
                 In the FM phase the overlap $q$ (green solid line) increases when cooling down
                 and splits into $q_0$ and $q_1$ (red lines) when the (continuous) transition to the
                 the FM$_{\rm 1RSB}$ phase is reached. The breaking parameter $x$ (blue line) is, 
                 discontinuous, and jumps from zero to a finite value.
                 The metastable FM$_{\rm 1RSB}$ solution eventually disappears  at $T_{\rm spFM}=0.4794$
                  and below this point no ferromagnetic solution exists. 
}}
\label{path}
\end{center}
\end{figure}

\Fref{potential} shows the Franz-Parisi potential for the spherical 3+4-FM spin glass model with 
$S^{\rm eff}_4=\beta_e/2$, i.e., the free energy of the configurations linked to that at the equilibrium 
temperature $T_e$, versus the magnetization $m$ for $T_e=0.805$, close to $T_d$, 
and different $T_a$.
The position on the cooling path is shown in the left panel of \Fref{path}, where the displayed
temperatures $T_a$ are indicated by points. The right panel of \Fref{path} shows the value of the
parameters $q, q_0, q_1, m, x$ along the cooling path  from $T_a=T_e=0.805$ down to
$T_a = T_{\rm spFM}=0.4794$.
For $T_a>T_s$, \Fref{potential} (a), the Franz-Parisi potential does
not have any secondary minimum at $m > 0$ besides the one at $m=0$. 
When $T_{\rm 1RSB}<T_a<T_s$, see  \Fref{potential} (b),  a secondary local minimum
at $m^{\star}>0$ develops in the RS potential. At $T_{\rm 1RSB}=0.7525$, see \Fref{potential} (c), 
this minimum becomes unstable towards 1RBS, the dashed part of the curves. When cooling
further, the minimum in RS free energy disappears, while it still exists for
the 1RSB  free energy, see \Fref{potential} (d). The latter  disappears
when the reentrance of the FM spinodal is reached for $T_a= T_{\rm spFM}=0.4794$, see \Fref{potential} (e), and beyond
this point the Franz-Parisi potential ceases  to have a minimum for 
$m>0$, see \Fref{potential} (f).
These results are equivalent to those of Ref.~\cite{BarratFranz97}, 
though there the case represented in  \Fref{potential} (f) was not considered.

\subsection{A loose end in the following states}
\label{KZ_instability}

Starting from the equations for the Langevin dynamics, authors of
Ref.~\cite{BarratFranz97} and Ref.~\cite{CaponeCastellani06} derive
explicitly the adiabatic evolution of order parameters for the
spherical 3+4-FM spin glass model, see Equation
(25)-(26)\footnote{There was a mistake in Equation (26) of Ref.~\cite
  {BarratFranz97} that has been corrected in
  Ref.~\cite{CaponeCastellani06}.} in \cite{BarratFranz97} or Equation
(12) in Ref.~\cite{CaponeCastellani06} (notice that there $J_4=0.45
J_3$ as chosen rather than $J_4=J_3$).  These equations are exact
description of the dynamics for $T_a>T_{\rm 1RSB}$ (denoted $T_{\rm
  ag}$ in Ref.~\cite{CaponeCastellani06}).

\begin{figure}[ht]
\begin{center}
\includegraphics[width=0.7\textwidth]{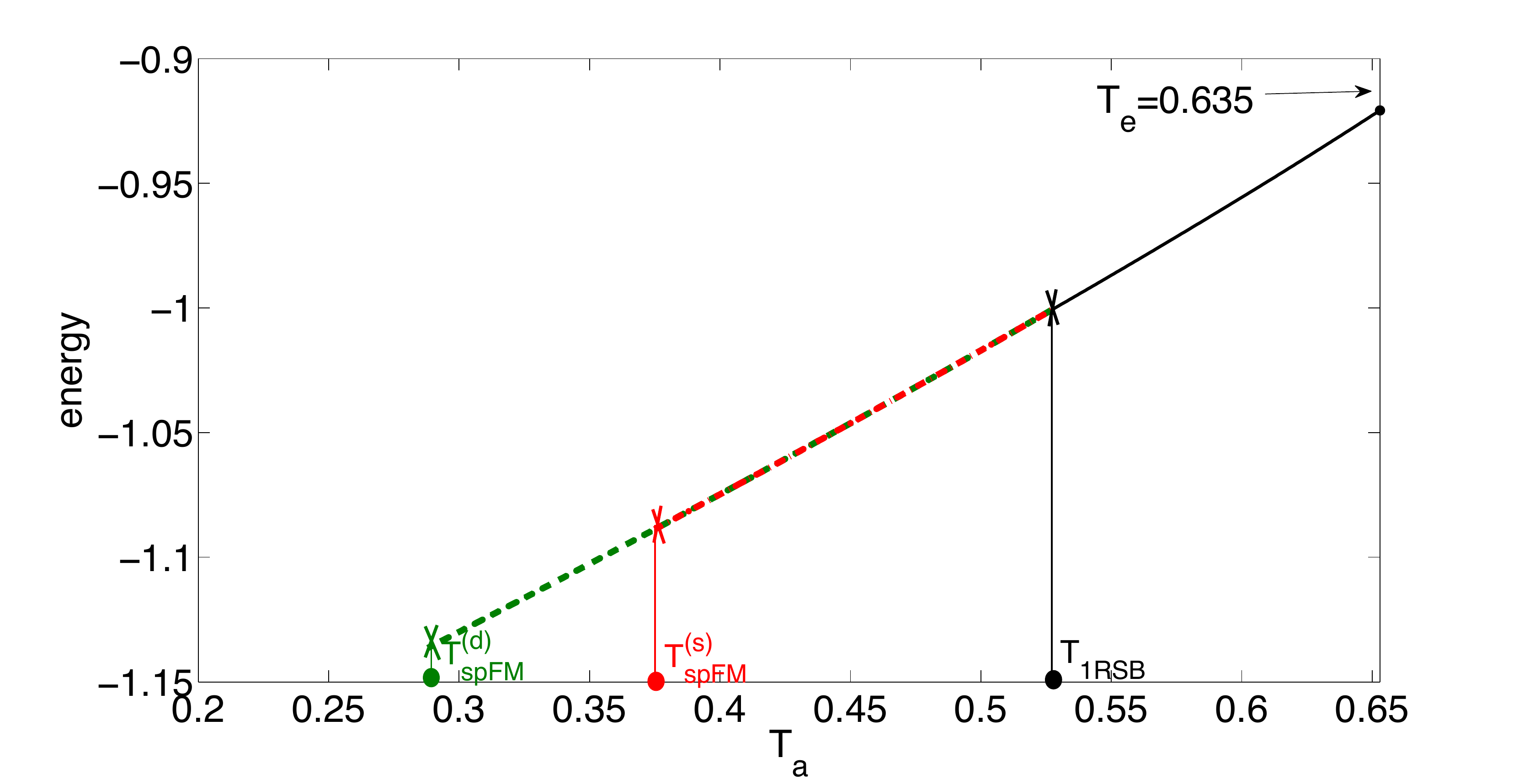}\\
\includegraphics[width=0.7\textwidth]{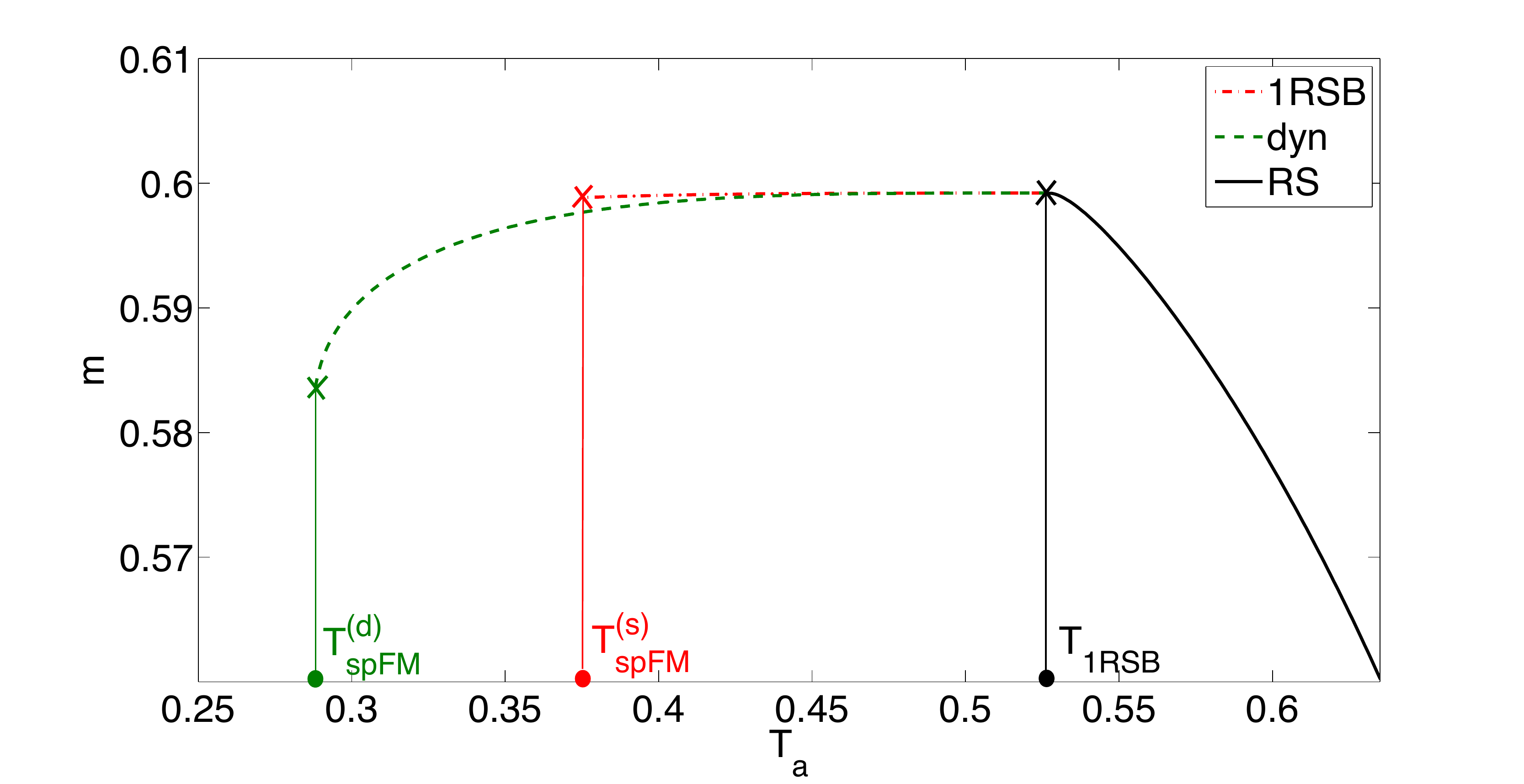}
 \caption{\small{
                 Energy (top) and magnetization $m$ (bottom) as function of $T_a$ along the path
                $T_e=0.653$ for the spherical 3+4-FM spin glass model 
                with $J_3=1$, $J_4=0.45$ (to compare with results of 
                 Ref.~\protect{\cite{CaponeCastellani06}}). 
                 At $T_a=T_{\rm 1RSB}$ the RS solution becomes unstable and the FM$_{\rm 1RSB}$ appears.
                 Red: results from the following states method,
                 Green: results from the approximate dynamic solution of Ref.
                 \protect{\cite{BarratFranz97,CaponeCastellani06}}.
                 The following of states ends at the $T_{\rm spFM}^{(s)} = 0.355$ and 
                 $T_{\rm spFM}^{(d)} = 0.288$, respectively, where the corresponding spinodal lines
                 are reached.
                 The energies of the two different method are almost indistinguishable, 
                 however but $E^{(s)}< E^{(d)}$. }
              }
\label{dynamic_solution}
\end{center}
\end{figure}

For $T_a<T_{\rm 1RSB}$ the dynamical solution of
\cite{BarratFranz97,CaponeCastellani06} is approximate because aging
appears within states for such temperatures. The dynamical equations
they obtain are the same as our 1RSB equations
(\ref{f:sp0})-(\ref{f:spm}), and also coincide with stationarity of
Franz-Parisi potential function with respect to $q_1$,$q_0$ and $m$.
The stationary condition (\ref{f:spx}) is, however, replaced by the
marginal condition (\ref{f:dyn_marg}), as expected from a dynamical
calculation.

We then conclude that also the dynamical solution provided in
Ref.~\cite{CaponeCastellani06} disappears because of the reentrance in
the phase diagram of the spinodal line in the spherical 3+4-FM spin
glass model, although at temperature $T_{\rm spFM}^{\rm (d)}$ slightly
lower than the reentrant spinodal $T^{\rm (s)}_{\rm spFM}$ found from
the static calculation.  We plot the corresponding energy in the top
panel of \Fref{dynamic_solution}.  The bottom panel
\Fref{dynamic_solution} shows the value of $m$ for the two solutions.
The vanishing states at low temperature was not noticed in Refs.
\cite{BarratFranz97, CaponeCastellani06}, because the data were for
higher temperatures.

The disappearing of the states at the spinodal line rises the question
of what may happen below this point.  In the model we have studied the
state, indeed, disappears. Therefore, the true long-time dynamics
would completely de-correlates from the initial configuration
($m=0$). We are unable, so far, to predict at which energy the
long-time dynamics will go from this point: this is a fundamental
limit of the following state approach. It would be interesting to see
if another approach allows to infer the long-time dynamics.

An interesting question is whether this behavior will be seen in other
models with discrete, rather than continuous, variables. In fact,
states vanishing seems to appear in many models, including the Ising
counter-part of the $p$-spin model. However at low temperature the
phase space of fully connected Ising models is more complex and is
described by a FRSB {\sl Ansatz}.  This opens the question of the
interplay between the state following method and the complexity of
phase space.  Hopefully, this can be studied along the line of the
spherical case here presented.
 
Another direction to study the fate of the states at
temperature $T_{\rm spFM}$ is to perform long and slow Monte-Carlo
simulations starting from an equilibrated configuration (this can be
achieve using the planting trick
\cite{KrzakalaZdeborova09,KrzakalaZdeborova11}). We view these
questions as an important open problem in the replica theory and
pointing this clearly out is one of the main aims of this article.

\section{Conclusion}
In this paper, we study the adiabatic evolution of states that are at
equilibrium at some temperature $T_e$ ($T_K \leq T_e \leq T_d$) in the
spherical 3+4-FM spin glass model.  We stress that while the study was
performed for the 3+4-FM spin glass model, the results are generic for
all $s$+$p$-FM spin glass model with only 1RSB low temperature phase.
 
After introducing the idea of planting an equilibrium configuration,
the following states problem is mapped to the physics of the
ferromagnetic solution in the corresponding model with a ferromagnetic
bias, i.e., the model along the Nishimori line. We exactly describe
the evolution of states away from equilibrium upon heating and
cooling.  This method is equivalent to the one based on the
Franz-Parisi potential \cite{BarratFranz97}.  Within our mapping the
Franz-Parisi potential is the free energy at fixed magnetization. Our
method also reproduces the known results about the long time dynamics
which is solvable for the spherical $p$-spin spin glass model
\cite{CaponeCastellani06}.

The most interesting outcome of this paper is identification of a sort
of boundary in the state following method and, hence, also in the
Franz-Parisi potential and the dynamical solution for the spherical
3+4-FM spin glass model. We show that it is related to a reentrant
behavior of spinodal lines in the phase diagram. For the states that
are at equilibrium close to the dynamic transition temperature, we
found that below a low but finite temperature, where we cross the
reentrant spinodal line, no solution with a finite magnetization
occurs, i.e., the correlation with the initial state at equilibrium is
completely, and discontinuously, lost.  In the spherical 3+4-FM spin
glass model no physical solution with more than one step RSB takes
place, thus no further fragmentation of states can be accounted to
remove the reentrance: the states simply stop to exist. This leaves
the following questions for future work: (1) Below the reentrance the
following state method, or equivalently the Franz-Parisi potential, is
unable to tell where and at which energy the dynamics will end. Is
there a way to perform a static computation answering this question?
(2) In the first part (RS) of the following state, the phase space is
relatively simple and decreasing $T_a$ just constraints the system
closer to the reference state and $m$ increases (in the ferromagnetic
notation). This lasts until we reach the FM/1RSB transition where the
phase space breaks down and becomes more complex. From this point on
it becomes harder to stay close the reference state because the
different states in which the phase space is broken into evolves
chaotically (chaos in temperature). What happens beyond the FM/1RSB
transition is thus an important point to study further. (3) Is the
behavior similar in the Ising models? While the RS and the 1RSB
solution both display a reentrance, other solutions breaking the
replica symmetry can exist and be self-consistent for discrete spin
systems at low temperature, so it is possible to conceive that the
reentrance shrinks and disappears allowing to follow states down to
zero temperature. This is again an important point to study further.

The answer to all those question is calling for more detailed
simulation of the Ising case, and more analytical studies. We hope
that this article will motive future attention and work in these
directions.

\section*{Appendix}
\label{RSB}
In this section, we recall the general condition that ensures whether
solutions with $R$-RSB ($R>0$) can exist. This proof has already been
presented, see \cite{CrisantiSommers91,CrisantiLeuzzi07}, and it is
reported here for completeness For any $R$ (including $\infty$) the
functional $G[\bm q, \bm m]$ \eref{f:G} can be written as \bea
\frac{1}{n}G[{\bm q},\bm m] = \int_{q_0}^{1}\!dq\,x(q)\,\Lambda(q) +
\int_{q_0}^{q_R}\!\frac{dq}{\chi(q)} + \ln\left(1 - q_R\right)
+\frac{q_0-m^2}{\chi(q_0)}+h(m)\nonumber \, , \nonumber
\\
\label{G:R-RSB}
\eea
where the function
\begin{equation}
x(q)=p_0+\sum_{r=0}^R(p_{r+1}-p_r)\theta (q-q_r)
\label{f:x_q}
\end{equation}
 is the cumulative probability density of the overlaps,
 $p_r$ ($p_0=n$ and $p_{R+1}=1$) are the sizes of the blocks along the diagonal and
 $q_r$ the value of $q_{ab}$ in the block,
 and
\begin{equation}
\chi(q) = \int_q^1\!dq'~x(q').
\label{f:chi_q}
\end{equation}
For what concerns Replica Symmetry Breaking, only the overlap variables are involved and not single replica index parameters, such as magnetization
$m_a$. Stationarity of the free energy functional with respect to $q_r$ and $p_r$ leads,
respectively, to the the self-consistency equations 
than can be concisely written as
\begin{eqnarray}
{\cal F}(q_r)=0 ,\qquad\qquad &\ &r=0,\ldots,R,
\label{f:self_q}
\\
\int_{q_{r-1}}^{q_r}dq~{\cal F}(q) = 0 ,
 \qquad&\ & r=1,\ldots,R,
\label{f:self_p}
\end{eqnarray}
where
\begin{equation}
{\cal F}(z)\equiv
\Lambda(z)-\int_0^{z}\!\frac{dq}{\chi(q)^2}.
\label{f:calF}
\end{equation}
Eq.~\eref{f:self_p} implies that ${\cal F}(q)$ has at least one
root in each interval $[q_{r-1},q_r]$, that, however,
 is not a solution of \eref{f:self_q}. 
Following Ref.~\cite{CrisantiLeuzzi07}, we, { then}, observe that
\eref{f:self_q}-\eref{f:self_p} guarantee that between any pair
$[q_{r-1},q_r]$ there must be at least two extremes
of ${\cal F}(q)$.
Denoting the extremes by $q^\star$, the condition ${\cal F}'(q^\star)=0$
leads to the equation, cf. Eq.~\eref{f:calF},
\begin{equation}
\chi(q^\star)\equiv\int_{q^\star}^1 x(q)~dq=\frac{1}{\sqrt{\Lambda'(q^\star)}},
\label{f:solutions}
\end{equation}
where
\begin{equation}
\Lambda'(q) = \frac{d \Lambda(q)}{dq}=\sum_p (p-1)\mu_p q^{p-2}.
\end{equation}
Since $x(q)$ is a non-decreasing function of $q$, cf.~\eref{f:x_q}, $\chi(q)$ is 
convex. The convexity of the function
$[\Lambda'(q)]^{-1/2}$ depends, instead, on the given values of multi-body interaction $g(p)$ considered, i.e., on the specific model, as well as on
the parameters $\mu_{p}$, i.e. on the phase diagram point of interest.
Non-zero values of the magnetization will affect the actual value of $q_0$,  it will be $q_0>0$ for
$m\not=0$, but it will not change the above argument since it only affects the values of $q$ at which  $\chi(q)$ displays steps, but not the convexity properties of the function.

\begin{figure}[t!]
\center
\includegraphics[width=.8\textwidth]{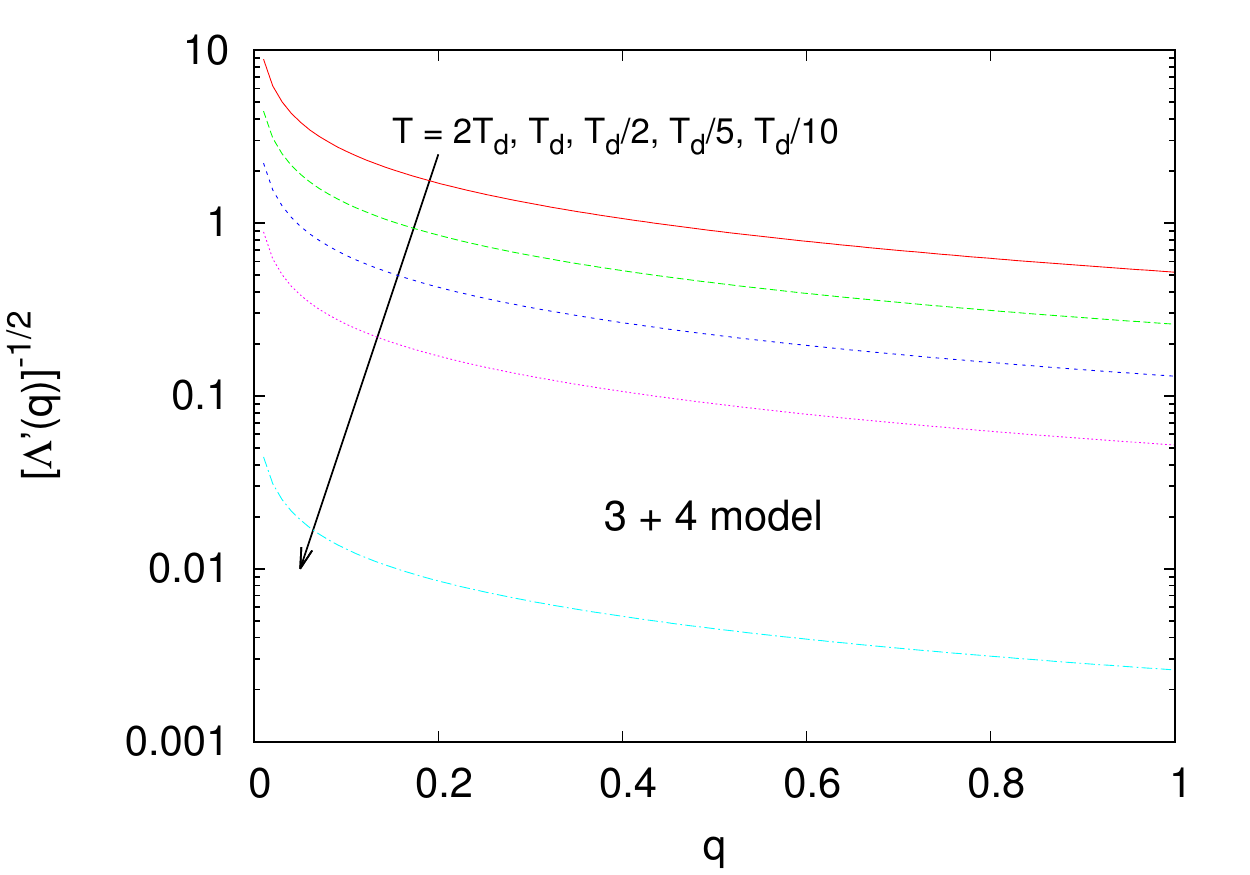}
\caption{\small Right hand side of \eref{f:solutions} for $J_3=J_4=1$ and $T=2T_d, T_d, T_d/2, T_d/5, T_d/10$, $T_d=0.805166$.
As explained in text,  the 3+4  curves never change convexity in any point of the phase diagram and can have no more than two
intersections with $\chi(q)$ ($q_0$ and $q_1$), implying at most a 1RSB solution.}
\label{fig:solutions}
\end{figure}
The $[\Lambda'(q)]^{-1/2}$ 
for the 3+4 model that we deal with throughout this paper, 
is plotted in \Fref{fig:solutions} 
in different points of the phase diagram.  
 The shape of
$[\Lambda'(q)]^{-1/2}$, concave, implies that at most a 1RSB solution can take
place, excluding, among others, also any solution with a continuous RSB.
This argument can be extended to show that 1RSB will be the most complicated solution for all $s+p$ 
systems in which $[\Lambda'(q)]^{-1/2}$  never becomes convex for $q\in [0:1]$. Given a value of $s$
 (equivalently of $p$) value this is quantitatively translated in satisfying the condition  
 $p<p^*(s)$ ($s<s^*(p)$), where $p^*$ (or $s^*$) is solution of
\begin{equation}
(p^2 + p + s^2 + s - 3 s p)^2 - p s (p-2) (s-2) = 0
\label{f:crit_L}\end{equation}
at fixed $s$ (or $p$), e.g., $(s^*,p^*)=(3,8)$, $(4,7+2\sqrt{6})$, $(5,9+3\sqrt{5})$.

\section*{References}
\bibliography{myentries}

\providecommand{\newblock}{}
\begin{thebibliography}{10}
\expandafter\ifx\csname url\endcsname\relax
  \def\url#1{{\tt #1}}\fi
\expandafter\ifx\csname urlprefix\endcsname\relax\def\urlprefix{URL }\fi
\providecommand{\eprint}[2][]{\url{#2}}

\bibitem{MezardParisi87b}
M{\'e}zard M, Parisi G and Virasoro M~A 1987 {\em Spin-Glass Theory and
  Beyond\/} ({\em Lecture Notes in Physics\/} vol~9) (Singapore: World
  Scientific)

\bibitem{FranzParisi95}
{Franz} S and {Parisi} G 1995 {\em Journal de Physique I\/} {\bf 5} 1401--1415

\bibitem{FranzParisi97}
Franz S and Parisi G 1997 {\em Phys. Rev. Lett.\/} {\bf 79} 2486--2489

\bibitem{KrzakalaZdeborova09b}
Krzakala F and Zdeborov\'a L 2010 {\em EPL\/} {\bf 90} 66002

\bibitem{ZdeborovaKrzakala10}
Zdeborov\'a L and Krzakala F 2010 {\em Phys. Rev. B\/} {\bf 81} 224205

\bibitem{CrisantiSommers91}
Crisanti A and Sommers H~J 1991 {\em Zeitschrift fur Physik B Condensed
  Matter\/} {\bf 87} 341--354

\bibitem{Nieuwenhuizen95}
Nieuwenhuizen T~M 1995 {\em Phys. Rev. Lett.\/} {\bf 74} 4289

\bibitem{CrisantiLeuzzi04}
Crisanti A and Leuzzi L 2004 {\em Phys. Rev. Lett.\/} {\bf 93} 217203

\bibitem{CrisantiLeuzzi07b}
Crisanti A and Leuzzi L 2007 {\em Phys. Rev. B\/} {\bf 76} 184417

\bibitem{CrisantiHornerSommers93}
Crisanti A, Horner H and Sommers H~J 1993 {\em Zeitschrift fur Physik B
  Condensed Matter\/} {\bf 92} 257

\bibitem{CugliandoloKurchan93}
Cugliandolo L~F and Kurchan J 1993 {\em Phys. Rev. Lett.\/} {\bf 71} 173

\bibitem{Barrat97}
Barrat A 1997 The p-spin spherical spin glass model cond-mat/9701031

\bibitem{CrisantiLeuzzi07}
Crisanti A and Leuzzi L 2007 {\em Phys. Rev. B\/} {\bf 75} 144301

\bibitem{GotSjo89}
G{\"o}tze W and Sj{\"o}rgen L 1989 {\em J. Phys.: Condens. Matter\/} {\bf 1}
  4203--422

\bibitem{CriCiu00}
Crisanti A and Ciuchi S 2000 {\em Europhys. Lett.\/} {\bf 49} 754--760

\bibitem{CriLeu12}
Crisanti A and Leuzzi L 2012 {\em in preparation\/}

\bibitem{Parisi80}
Parisi G 1980 {\em J. Phys. A Lett.\/} {\bf 13} L115--L121

\bibitem{CrisantiLeuzzi03}
Crisanti A, Leuzzi L and Rizzo T 2003 {\em Eur. Phys. J. B\/} {\bf 36} 129

\bibitem{CrisantiLeuzzi06}
Crisanti A and Leuzzi L 2006 {\em Phys. Rev. B\/} {\bf 73} 014412

\bibitem{Crisanti08}
Crisanti A 2008 {\em Nucl. Phys. B\/} {\bf 796} 425

\bibitem{KrzakalaZdeborova09}
Krzakala F and Zdeborov\'a L 2009 {\em Phys. Rev. Lett.\/} {\bf 102} 238701

\bibitem{KrzakalaZdeborova11}
Krzakala F and Zdeborov\'a L 2011 {\em J. Chem. Phys.\/} {\bf 134} 034513

\bibitem{MontanariRicci04}
Montanari A and Ricci-Tersenghi F 2004 {\em Phys. Rev. B\/} {\bf 70} 134406

\bibitem{BarratFranz97}
Barrat A, Franz S and Parisi G 1997 {\em J. Phys. A\/} {\bf 30} 5593--5612

\bibitem{CaponeCastellani06}
Capone B, Castellani T, Giardina I and Ricci-Tersenghi F 2006 {\em Physical
  Review B\/} {\bf 74} 144301

\bibitem{Gardner85}
Gardner E 1985 {\em Nuclear Physics B\/} {\bf 257} 747--765

\end{thebibliography}

\end{document}